\documentclass[11pt]{article}
\pdfoutput=1
\usepackage{jheppub,setspace}
\newcommand{\EQ}[1]{\begin{align}\begin{split} #1 
\end{split}\end{align}}
\newcommand{\eq}[1]{\begin{equation}{ #1 
}\end{equation}}
\def\CW{{\cal W}}
\def\a{\alpha}\def\b{\beta}\def\d{\delta}\def\D{\Delta}\def\g{\gamma}\def\G{\Gamma}\def\e{\epsilon}\def\z{\zeta}
\def\l{\lambda}\def\m{\mu}\def\s{\sigma}\def\t{\tau}

\begin{document}
\title{\Large Solution of quantum integrable systems from quiver gauge theories}
\author[1]{Nick Dorey} 
\author[2]{and Peng Zhao}

\affiliation[1]{DAMTP, University of Cambridge, Cambridge CB3 0WA, UK}
\affiliation[2]{Simons Center for Geometry and Physics, Stony Brook, NY 11794, USA}

\abstract
{
We construct new integrable systems describing particles with internal spin from four-dimensional $\mathcal{N}=2$ quiver gauge theories. The models can be quantized and solved exactly using the quantum inverse scattering method and also 
using the Bethe/Gauge correspondence. 
}
\maketitle
\section{Introduction}

The mysterious connections between integrable systems and supersymmetric gauge theories have lead to a fruitful interplay between the two subjects. One of the best-known examples is the relationship
between classical integrable systems and four-dimensional $\mathcal{N} = 2$
supersymmetric theories \cite{Gorsky:1995zq, Martinec:1995by, Donagi:1995cf}. The Seiberg-Witten curve encoding the low-energy dynamics of the gauge theory coincides with the spectral curve
encoding the mutually-commuting Hamiltonians of the integrable system. This coincidence has far-reaching consequences. Most importantly, it opens the
door for studying the long-standing problem of the quantization of the
Seiberg-Witten solution from the quantization of the corresponding integrable system, and vice versa. 

In the last few years, this connection has been made more precise by
the work of Nekrasov and Shatashvili \cite{Nekrasov:2009uh,
  Nekrasov:2009ui, Nekrasov:2009rc, Nekrasov:2014xaa}. They observed
that quantization is related to deforming the theory by an
$\Omega$-background in a two-dimensional plane. The rotation parameter
$\e$ in this plane is identified with Planck's constant $\hbar$. The
supersymmetric vacua of the $\mathcal{N}=2$ theory are in one-to-one
correspondence with the eigenstates of the quantum integrable system
labeled by the solutions of the Bethe ansatz. The so-called
Bethe/Gauge correspondence has given us new insights into dualities and symmetries between gauge theories \cite{Orlando:2010aj, Hellerman:2011mv, Orlando:2011nc, 
  Hellerman:2012zf, Muneyuki:2011qu, Mironov:2012uh, Gadde:2013wq, 
  Bulycheva:2012ct, Chen:2013jtk, Luo:2013nxa, Bonelli:2014iza, Bonelli:2015kpa,
  Koroteev:2015dja, Bourgine:2015szm}.\footnote{See \cite{Lamers:2012thesis} for a
  pedagogical introduction to the Bethe/Gauge correspondence.} For
example, it has been used to establish new 2d/4d dualities
\cite{Dorey:2011pa, Chen:2011sj} and 3d/5d dualities
\cite{Chen:2012we, Nieri:2013vba, Bullimore:2014awa}, to shed light on
2d Seiberg-like dualities \cite{Orlando:2010uu,
  Benini:2014mia}, and 3d mirror symmetry
\cite{Gaiotto:2013bwa}. The correspondence also solves quantum
integrable models in finite volume, as it gives rise to
thermodynamic Bethe ansatz equations by summing the 
instantons \cite{Nekrasov:2009rc, kozlowski2010tba, 
Meneghelli:2013tia, Bourgine:2014yha, Hatsuda:2015qzx}.

 Moreover, it has been conjectured that the supersymmetric vacua of
 \emph{any} $\mathcal{N}=2$ theory in the Nekrasov-Shatashvili
 background corresponds to the solution of a quantum integrable
 system. Therefore finding the gauge-theoretic ``dual" of a given
 classical integrable system will establish integrability at the
 quantum level. In the other direction, systematically identifying the
 integrable model ``dual" to a given gauge theory 
is an intriguing open problem.

In this paper, we use the Bethe/Gauge correspondence to quantize and
solve a new class of 
integrable systems arising from 4d $\mathcal{N}=2$
elliptic quiver gauge theories.\footnote{The Seiberg-Witten geometry of this class of quiver gauge theories have recently been studied in \cite{Nekrasov:2012xe,
Nekrasov:2013xda}.} The Coulomb branches of the gauge theories can be
described as algebraic integrable systems with commuting Hamiltonians
parametrized by a set of holomorphic coordinates constructed from the hyper-K\"ahler quotient. Real integrable systems arise on taking an appropriate middle-dimensional real section of the complex phase space. A very general class of integrable systems can be engineered this way describing particles with internal degrees of freedom. This class contains many well-known integrable systems such as the Calogero-Moser model and the Heisenberg spin chain in special corners of the parameter space. 

Here we will consider two models, one of which is well-known. 
Both correspond to systems of $K$ particles moving in one dimension subject to periodic boundary conditions. We will denote the (real) positions and conjugate momenta of the particles as $x_{k}$ and $p_{k}$ respectively, $k=1,\ldots,K$. Each
particle carries internal degrees of freedom corresponding to $N$ harmonic oscillators. 
For the $k$-th particle we have annihilation and creation operators $Q^{\alpha}_{k}$, $\tilde Q^{\, \alpha}_{k}$ with $\alpha=1,\ldots,N$. 
Both models have an internal symmetry group of rank $N-1$ corresponding to this index. In the classical version of each model, the variables described above obey canonical Poisson brackets. Using standard techniques from the theory of integrable systems we will construct quantum systems in which the corresponding operators obey canonical commutation relations.  

The first model we consider is the elliptic spin Calogero-Moser model. In this case the particles carry classical $\mathfrak{sl}(N)$ ``spins''
which are constructed from the oscillators in the standard way,  
\eq{
S^{~\!\!\a\b}_{k} = Q^{\a}_{k}\tilde Q^{\b}_{k} - \frac{\d^{\a\b}}{N}\sum_{\g=1}^N Q^{\g}_{k}\tilde Q^{\g}_{k}. 
}
The classical Hamiltonian is given as,   
\eq{
H = \sum_{k=1}^{K} \frac{p_{k}^{2}}{2} \,\,+\,\, \sum^K_{\ell > k}\sum_{\a,\b=1}^N\,
S^{~\!\!\a\b}_{k}S^{~\!\!\b\a}_{\ell}\, \wp(x_{k}-x_{\ell}), 
}
where $\wp(z)$ is the Weierstra$\ss$ elliptic function
defined on a torus of complex structure $\t$. The periodicity of
this function for real arguments yields a system of particles moving
in a box of size $L\sim \text{Im}~\!\t$ subject to periodic boundary
conditions. 
 
As we review below, the classical model arises as a
particular real section of the Coulomb branch of an $\hat{A}_{N-1}$
quiver gauge theory with gauge group $G=U(1)\times SU(K)^{N}$. 
The parameter $\t$ corresponds to the complexified gauge coupling of
the diagonal $U(K)$ subgroup of $G$. The off-diagonal gauge couplings
are tuned to a particular strong-coupling point where a hidden global
$A_{N-1}$ symmetry appears. Following the recipe introduced by
Nekrasov and Shatashvili, quantization is achieved by introducing an
$\Omega$-background in one plane. The induced twisted superpotential
of the resulting 2d effective theory corresponds to the Yang-Yang
potential which determines the spectrum of the corresponding 
quantum integrable system. To select the real section corresponding to
the spin Calogero-Moser model, it is also necessary to choose an
appropriate electro-magnetic duality frame for the quiver gauge
theory. This point is discussed further in section \ref{section4} below.   
 
In principle, with the above identification, the
Nekrasov-Shatashvili procedure provides a quantization of the model
for all values of the parameters. Here, we will focus on the 
large-volume limit $L\sim\text{Im}~\!\t \gg 1$, where the system can 
also be solved using the {\em asymptotic Bethe ansatz}. The idea of
the asymptotic Bethe ansatz is to first solve the problem in the limiting case 
$\text{Im}\,\t=\infty$ where the particles move on an infinite line,
with the $k$-th and the $\ell$-th particles 
interacting via the two-body potential, 
\eq{
V(x_{k}-x_{\ell}) \sim \sum_{\a,\b=1}^N\frac{S^{~\!\!\a\b}_{k}S^{~\!\!\b\a}_{\ell}}{4\sinh^{2}\left(\frac{x_{k}-x_{\ell}}{2}\right)}. 
}
This gives rise to a scattering problem for asymptotic states
corresponding to free particles carrying classical spins
$S^{~\!\!\a\b}_{k}$. For these asymptotic states, quantization
proceeds in a straightforward way by promoting the canonical Poisson
brackets of the variables
$\{x_{k},\,p_{k},\,Q^{\a}_{k},\, \tilde Q^{\a}_{k}\}$ to
canonical commutation relations. For appropriate values of the
conserved quantities, the
resulting spin operators $\hat{S}^{~\!\!\a\b}_{k}$ act in 
lowest-weight irreducible representations of
$\mathfrak{sl}(N,\mathbb{R})$. At least for $L\sim \text{Im}~\!\t \gg 1$,
the quantum model can be thought of as a system of $K$ particles each carrying
a non-compact ``spin'' corresponding to a lowest-weight representation of $\mathfrak{sl}(N,\mathbb{R})$.  
 
Quantum integrability of the model requires that multi-particle
scattering factorizes into a product of successive two-body scattering
processes. Furthermore, the consistency of factorized scattering 
requires that the two-body S-matrix obey the Yang-Baxter equation. Our
approach here, will be to {\em assume}  factorization of
multi-particle scattering. However, we will check the Yang-Baxter
equation explicitly. The first step in the analysis is to solve the 
Schr\"{o}dinger equation describing the scattering of two of these
particles. As advertised, the resulting two-body S-matrix indeed obeys
the Yang-Baxter equation. Through our assumption of factorization, the
multi-particle S-matrix is then determined. We find that it can be diagonalized
explicitly using the quantum inverse scattering method. The last step is to
impose periodic boundary conditions on the resulting scattering
wave functions which leads to the asymptotic Bethe ansatz equations. The energy spectrum of the model is
then determined by solutions of these equations. Our main result is
that the Nekrasov-Shatashvili quantization procedure applied to the quiver gauge theory, yields the same Bethe ansatz equations and therefore the same spectrum.   

The second model we study involves a different limit of the parameters
of the full inhomogeneous system. For the original elliptic
Calogero-Moser model for $K$ particles without spin, with Hamiltonian, 
\eq{
H =  \sum_{k=1}^{K} \frac{p_{k}^{2}}{2} \,\,+\,\, m\sum^K_{\ell> k}\,\wp(x_{k}-x_{\ell}). 
}
There is a well-known limit, first discussed by Inozemtsev \cite{Inozemtsev1989finite}, which yields the $K$-body Toda
chain with Hamiltonian,
\eq{
H^\text{Toda} = \sum_{k=1}^{K}\frac{p^{2}_{k}}{2} + \sum_{k=1}^{K-1}e^{X_{k}-X_{k+1}} + \Lambda^{2K}e^{X_{K}-X_{1}\label{TodaHamiltonian}},
}
where $\Lambda=m\exp(2\pi i\t/K)$. 
In the classical version of the correspondence to supersymmetric gauge
theory, the scalar elliptic model corresponds to the ${\cal N}=2$ super 
Yang-Mills theory with an adjoint hypermultiplet of mass $m$ and complexified
coupling $\t$ (also known as the ${\cal N}=2^*$ theory). The
Inozemtsev limit coincides with the standard decoupling limit for the adjoint hypermultiplet which yields
the minimal ${\cal N}=2$ gauge theory. The latter is asymptotically
free and is characterized by the RG-invariant scale $\Lambda=m\exp(2\pi i\t/K)$. 

Here, we will take a similar limit for the elliptic quiver gauge theory which yields a
Toda-like chain for particles with internal degrees of freedom.   
As before we have $K$ particles moving in one dimension 
with positions $x_{k}$ and momenta $p_{k}$, $k=1,\ldots,K$, each
particle having $N$ internal harmonic oscillator degrees of freedom 
with annihilation and creation operators $Q^{\a}_{k}$, $\tilde Q^{\a}_{k}$ for $\a=1,\ldots,N$. Now we form $\mathfrak{sl}(N)$-invariant hopping operators between the $k$-th and the $\ell$-th sites, 
\eq{
A_{k\ell} = \sum_{\alpha=1}^{N} Q^{\a}_{k}\tilde Q^{\a}_{\ell}. \label{hopping}
}
By taking an Inozemtsev-like limit, we find a classical integrable
system with quadratic Hamiltonian, 
\eq{
H^\text{HT} = H^\text{Toda} + \sum^{K}_{k=1}\frac{A_{kk}^{2}}{4} +
\frac{1}{2}\Bigg[\sum^{K-1}_{k=1}e^{\frac{X_{k}-X_{k+1}}{2}}\left(A_{(k+1)k}
  + A_{k(k+1)} \right) +
\Lambda^{K}e^{\frac{X_{K}-X_{1}}{2}}\left(A_{1K} + A_{K1}
\right)\Bigg].
\label{HTa}
}
We check directly the classical integrability of this model. 
 
As above we study the quantization of the above system in the
framework of the asymptotic Bethe ansatz, which gives an accurate
description of the system in the limit of large volume. In this case,
the quantum system consists of $K$ particles, interacting via
exponential potentials, each carrying $N$
harmonic oscillator degrees of freedom. The corresponding occupation
numbers are individually conserved when the particles are far
apart. However, the interaction terms in the Hamiltonian 
proportional to $A_{k (k+1)}$, mean that occupation number can be
transferred from one particle to the next one in the chain. Freezing 
the positions of the $K$ particles, the resulting dynamics of the
oscillator degrees of freedom is closely related to the Hubbard
model. For this reason we propose to call the system (\ref{HTa}), the
{\em Hubbard-Toda chain}. Once again our main result is a comparison
of the large-volume solution of the model via the asymptotic Bethe
ansatz with the appropriate application of the Bethe/Gauge
correspondence, which yields exact agreement. 

The Bethe/Gauge correspondence not only provides a quantization of the
corresponding classical integrable system, but also provides the
solution to the full quantum problem. Since the Bethe ansatz equations
are mapped directly to the supersymmetric vacua of the quiver gauge
theory, the vacuum equations provide a prediction for the scalar part
of the S-matrix. The prediction agrees perfectly with the direct
solutions of the matrix Schr\"odinger equation. Furthermore, the
instanton partition function yields a set of thermodynamic Bethe
ansatz equations that determine the finite-size spectrum of the model.

This paper is organized as follows. In section \ref{section2}, we describe the
brane setup of the elliptic quiver and 
how classical integrable systems arise from the Coulomb and the Higgs branch descriptions. 
In section \ref{section3}, we quantize the integrable system and exactly solve the
system using the quantum inverse scattering method. In section \ref{section4}, we
use the Bethe/Gauge correspondence to predict the scalar part of the
S-matrix. The appendices contain details on the Inozemtsev limit and
the classical integrability of the Hubbard-Toda model.

\section{Integrable systems from elliptic quiver gauge theories}\label{section2}
\subsection{The brane setup}
We consider 4d $\mathcal{N}=2$ quiver gauge theories whose quiver diagram is the affine Dynkin diagram of $\hat A_{N-1}$ type. The gauge group is $U(1)_D \times SU(K)^N$. There is a vector multiplet for each $SU(K)$ factor and a bi-fundamental hypermultiplet for adjacent $SU(K)$ factors. Each gauge group factor has a gauge coupling $g_\a$. The $\b$ function vanishes and the theories are conformal. We combine the gauge coupling and the theta angle into a marginal gauge coupling $\t_\a = 4\pi i/g^2_\a + \vartheta_\a/2\pi$. The theories at low energy have a moduli space of vacua known as the Coulomb and the Higgs branches. In the Coulomb branch, the complex scalars in the vector multiplet acquire vacuum expectation values and the gauge group is broken down to its Cartan subgroup. In the Higgs branch, the complex scalars in the hypermultiplet acquire vacuum expectation values and break the gauge group completely.

The quiver gauge theories can be embedded in string theory as the world-volume theories of $K$ D4-branes intersecting $N$ NS5-branes in the Type IIA string theory. The D4-branes have world volume in the $01236$ direction and is compactified in the $x^{6}$ direction. The NS5-branes have world volume in the $012345$ direction. This setup is called the elliptic model as it arises from M-theory, which has an additional compact $x^{10}$ direction \cite{Witten:1997sc}. We will refer to them as elliptic quiver theories to distinguish from the corresponding elliptic integrable models.

\begin{figure}[h]
\center
\includegraphics[scale=0.7]{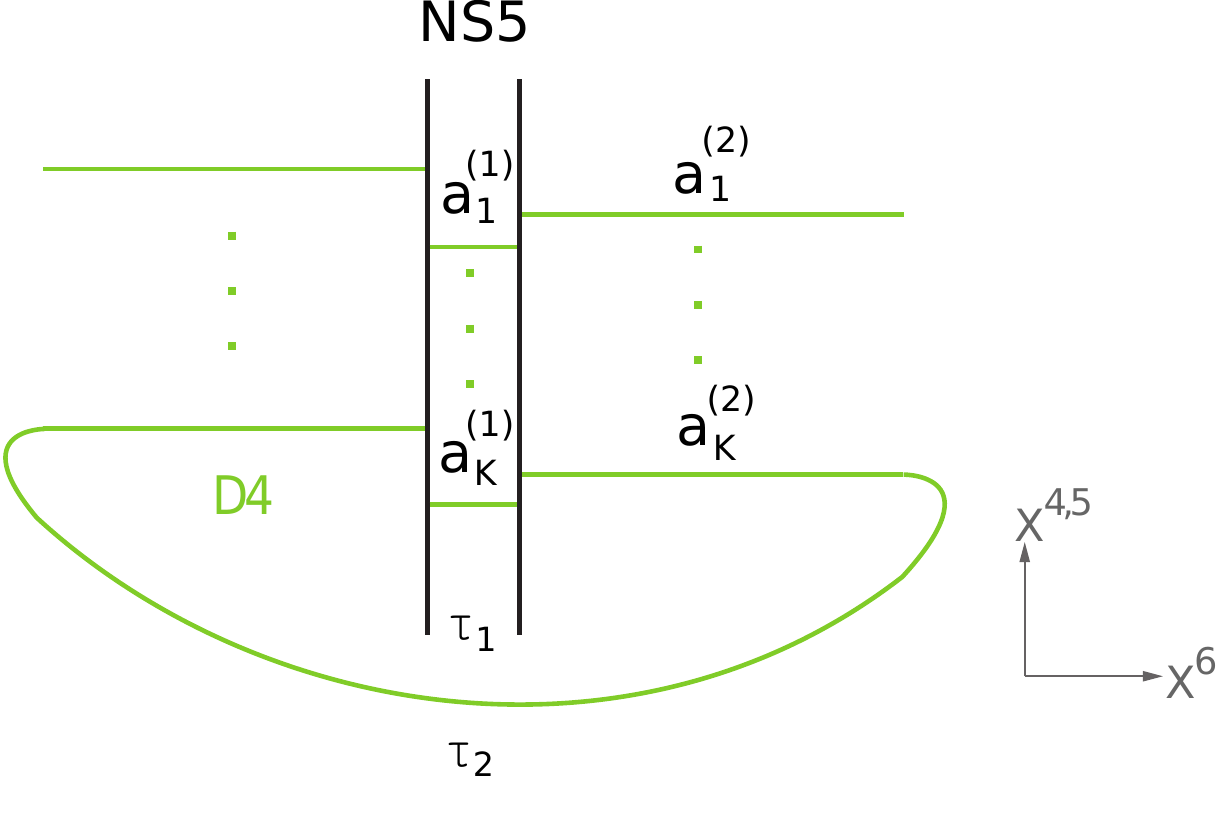}
\caption{The Type IIA brane construction of the $U(1)_D \times SU(K)^2$ quiver gauge theory.}
\label{brane}
\end{figure}
The generic brane configuration is shown in figure \ref{brane}. The brane setup preserves eight real supercharges and engineers a 4d $\mathcal{N}=2$ gauge theory with gauge group $U(1)_D \times SU(K)^{N}$. Throughout the paper, we will use $\a, \b = 1, \ldots, N$ to denote $SL(N)$ indices, and $k, \ell = 1, \ldots, K$ to denote $SU(K)$ indices. The positions $(a^{(\a)}_{1}, \ldots, a^{(\a)}_{K})$ of the D4-branes between the $\a$-th and the $(\a+1)$-th NS5-branes label the Coulomb branch moduli of the $\a$-th  $SU(K)$ factor of the gauge group. The diagonal $U(1)_D$ factor corresponds to the center-of-mass position of all the D4-branes and decouples from the low-energy dynamics. The relative center-of-mass positions between the neighboring D4-branes define the mass of the bi-fundamental hypermultiplets $m_{\a}$. The mass can be arbitrarily chosen by imposing a twisted periodicity condition on the $x^{6}$ circle: $a^{(\a+N)}_{k} = a^{(\a)}_{k} + m$ as $x^{6} \to x^{6} + 2\pi R_{~\!\!6}$ such that $\sum_\a m_\a = m$. Because we consider an equal number of D4-branes on either side of the NS5-brane, the theory is conformal. The separation of the NS5-branes in the $x^{6}$ and the $x^{10}$ directions are proportional to the gauge coupling $1/g^2_\a$ and the theta angle $\vartheta_\a$, respectively. The gauge coupling $1/g^2 = \sum_\a 1/g^2_\a$ of the diagonal subgroup $U(K) = U(1)_D \times SU(K)/\mathbb{Z}_K$ is proportional to the radius $R_{~\!\!6}$ of the $x^6$ circle.

In the limit when one of the gauge couplings becomes weakly coupled, the corresponding gauge group can be frozen to become a global symmetry. The dynamical D4-branes parametrizing the Coulomb branch moduli become rigid D4-branes labeling the flavor charges. The elliptic quiver then reduces to a linear quiver. 
Conversely, the elliptic quiver can be obtained from the linear quiver by weakly gauging the global symmetry.  As we will see, these have clear analogues on the integrable systems side where the weak-coupling limit corresponds to taking the infinite-volume limit. The two-body S-matrix is well-defined and can be solved for on the infinite line. We then pass to a large circle and use the asymptotic Bethe ansatz to determine the spectrum of the system.

\begin{figure}[h]
\center
\includegraphics[scale=0.6]{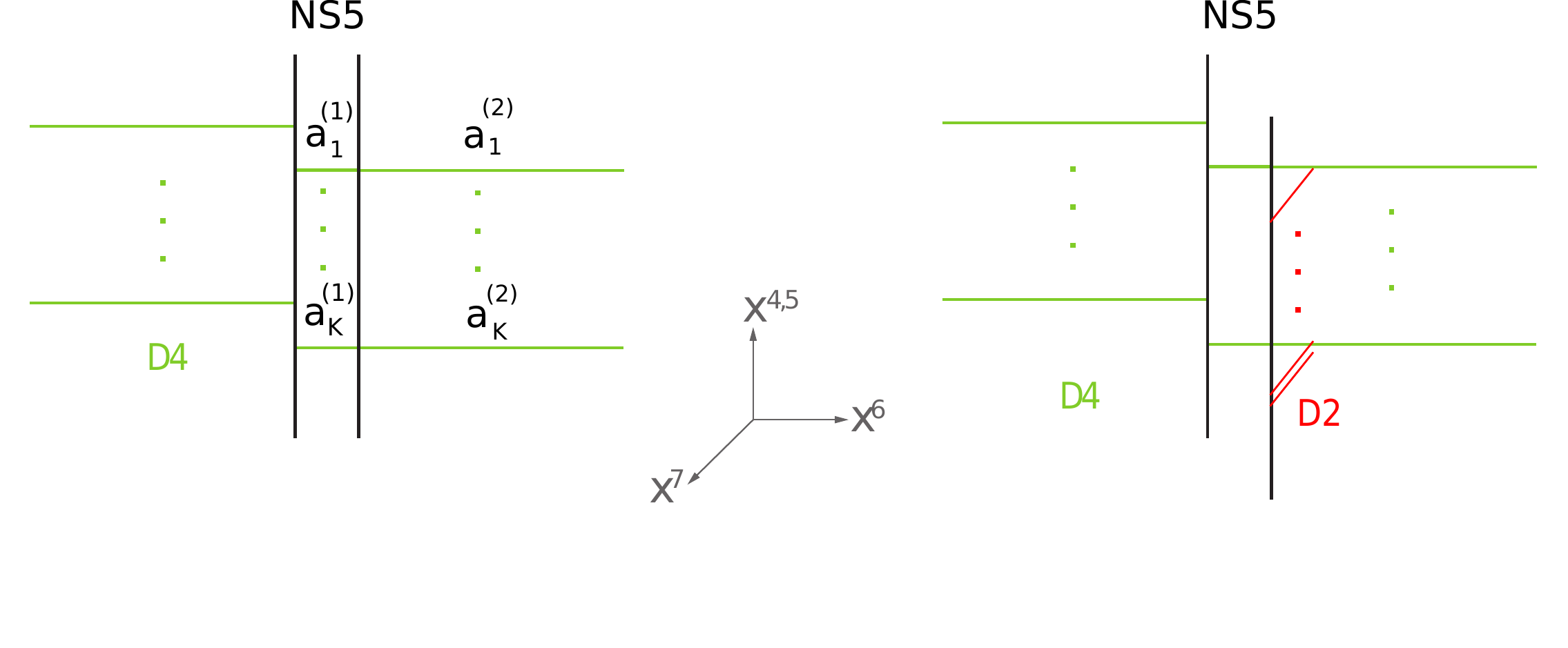}
\caption{At the Higgs branch root, D4-branes on either side of an NS5-brane reconnect. We move onto the Higgs branch by lifting the NS5-brane in the $x^{7}$ direction. The coupled 2d-4d system describes vortex strings probing the 4d theory.} 
\label{brane1}
\end{figure}

There is a special point in the Coulomb branch moduli space where $a^{(\a)}_{k} = a^{(\a+1)}_{k}$ for all $k$ and for $\a=1, \ldots, N-1$. At this point, the D4-branes on either side of the $N-1$ NS5-branes coincide and reconnect. The $N-1$ NS5-branes can then be lifted in the orthogonal $x^{7}$ direction and the 4d theory moves onto its Higgs branch. 
The special point at which the Coulomb and the Higgs branch meet is called the Higgs branch root, as shown on the left of figure \ref{brane1}. The separation of the NS5-branes in the $x^7$ direction corresponds to the Higgs branch vacuum expectations value.  The theory in the Higgs branch admits vortex string solutions. They appear as D2-branes stretched between the lifted NS5-branes and the D4-branes in the $0127$ direction \cite{Hanany:2003hp, Hanany:2004ea}, as shown on the right of figure \ref{brane1}. The vortex string tension is proportional to the Higgs branch vacuum expectation value. The number of D2-branes $M_{\a}$ is arbitrary. The world-volume theory on the D2-branes is a 2d $\mathcal{N}=(2, 2)$ gauged linear sigma model with gauge group $U(M_{1}) \times U(M_{1}+M_{2})\times \cdots \times U(M_{1} + \cdots + M_{N-1})$ \cite{Hanany:1997vm}. The fundamental strings stretched between the D2 and the D4-branes define the fundamental and the anti-fundamental chiral multiplets. The separations of the NS5-branes in the $x^6$ direction is the Fayet-Iliopoulos parameter $r_\a$ of the 2d theory. It combines with the 2d theta angle to form the complexified coupling $\hat \t_\a = ir_\a + \theta_\a/2\pi$, which will be identified with the 4d gauge coupling $\t_\a$.
 When the Fayet-Iliopoulos parameter is turned off, the 2d theory is in its Coulomb branch parametrized by the vacuum expectation values of the twisted chiral multiplet scalars $\s^{(\a)}_i$, which label the positions of the D2-branes in the $x^4+ix^5$ plane.

Surprisingly, the 2d theory captures the physics of the 4d theory. This was first suggested by matching the BPS spectra of the two theories \cite{Dorey:1998yh, Dorey:1999zk}. This 2d/4d duality was made more precise when the 4d theory is subject to the $\Omega$-background in the Nekrasov-Shatashvili limit \cite{Dorey:2011pa, Chen:2011sj}.  In this deformed background, the 4d theory is localized onto a 2d subspace preserving $\mathcal{N}=(2, 2)$ supersymmetry. The theory is described by an effective twisted superpotential, which coincides with that of the theory living on the vortex string.

We will be interested in two special configurations and their decoupling limits.
\begin{enumerate}
\item We take the limit that the diagonal $U(1)_D \times SU(K)$ becomes weakly coupled with gauge couplings $1/g_\alpha$, $\alpha = 1,\ldots, N-1$ held fixed. This implies that $g \to 0$ and hence $g_N \to 0$. In this limit, the theory factorizes into a diagonal $U(1)_D \times SU(K)$ gauge group and a linear quiver with gauge group $SU(K)^{N-1}$ . The diagonal gauge group has an adjoint hypermultiplet of mass $m$, while the linear quiver has $K$ fundamental hypermultiplets and $K$ anti-fundamental hypermultiplets. The Coulomb branch moduli of the diagonal gauge group appear as mass parameters for the  hypermultiplets in the linear quiver. The non-trivial periodicity condition defines the adjoint hypermultiplet mass $m$. For the $U(1)_D \times SU(K)^2$ theory shown in figure \ref{brane}, this corresponds to taking $\text{Im}\, \t_1 \to 0$ while sending $\text{Im}\,\tau_{~\!\!2} \to \infty$. In terms of integrable systems, this is the large-volume limit of the elliptic spin Calogero-Moser model. 
\item As before we take the diagonal coupling to zero but now we send $g_1  = g_N \to 0$ such that the gauge group factorizes into a weakly-coupled $SU(K) \times SU(K)$ and an $SU(K)^{N-2}$ linear quiver. The $SU(K) \times SU(K)$ has a bi-fundamental hypermultiplet of mass $\m = m_N$ and a bi-fundamental hypermultiplet of mass $\sum_{\a=1}^{N-1} m_\a$. The Coulomb branch moduli of the $SU(K) \times SU(K)$ gauge group appear as mass parameters for the fundamental and the anti-fundamental hypermultiplets. We take the limit where $\m$ becomes infinitely massive while the combination $\Lambda^{K} = \m^{K} e^{2\pi i \tau_{N}}$ is fixed. For a single gauge group factor, this corresponds to flowing from an $\mathcal{N}=2^*$ theory to a pure $\mathcal{N}=2$ theory by decoupling the adjoint hypermultiplet while taking the weak-coupling limit such that a dynamical scale $\Lambda$ is generated via dimensional transmutation.

In the brane picture, we take a single NS5-brane to be at half-period. 
This is depicted in the left of figure \ref{Inozemtsev}. In our limit, the half-period $i\pi\t_{N}$ and the separation $\m$ of the D4-branes ending on the single NS5-brane are sent to infinity. The NS5-brane effectively becomes two disjoint NS5-branes separated by a distance $\log \Lambda$ and each sourcing $K$ semi-infinite D4-branes ending on the stack of NS5-branes, as shown on the right of figure \ref{Inozemtsev}. As seen from the other $N-1$ NS5-branes, the two ends of the D4-branes are frozen and define a global symmetry group. The positions of these D4-branes define the mass of the fundamental and the anti-fundamental hypermultiplets $m_k$ and $\tilde m_k$. The theory effectively reduces to a linear quiver gauge theory.
\begin{figure}[h]
\center
\includegraphics[scale=0.6]{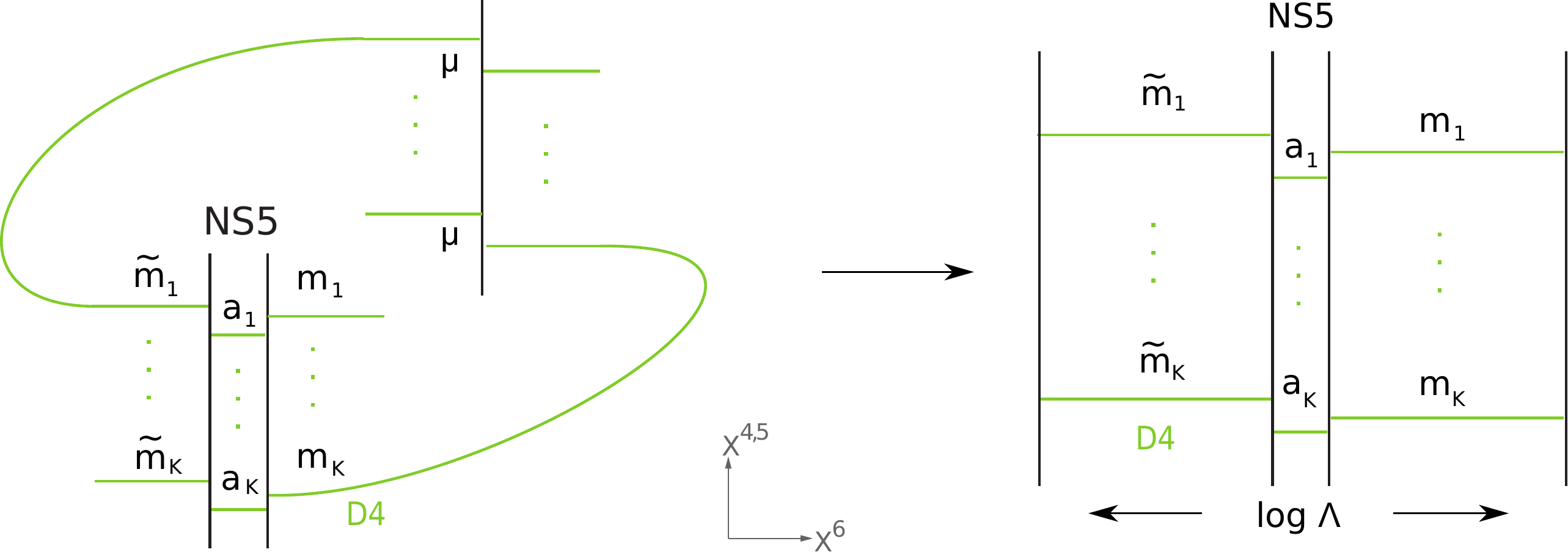}
\caption{Flowing from an $\mathcal{N}=2^*$ theory to a pure $\mathcal{N}=2$ theory: we take the weak-coupling limit and keep the combination $\Lambda^{K} = \m^{K}e^{2\pi i\t_N}$ fixed.}
\label{Inozemtsev}
\end{figure}
\end{enumerate}
Each example corresponds to a classical integrable system. The Seiberg-Witten curve coincides with the spectral curve of each integrable system. The chiral ring corresponds to the conserved Hamiltonians. By matching the curves, we may identify the parameters. 
We summarize the dictionary between gauge theories and integrable systems in table \ref{T1}.
\begin{table}[htbp]
\begin{center}
    \begin{tabular}{c|c}
        \hline
       Gauge theory & Integrable system \\ \hline\hline 
        $N$ NS5-branes & $\mathfrak{sl}(N, \mathbb{C})$ spin\\  \hline 
        $K$ D4-branes & $K$ particles \\ \hline
        $M$ D2-branes & $M$ magnons \\ \hline 
        Marginal coupling $\t$ & System size $L=\text{Im}~\!\t$ \\ \hline 
        Coulomb branch moduli $a_{k}$ & Particle momentum $p_{k}$ \\ \hline 
        Twisted chiral multiplet scalar $\s_{i}$ & Magnon rapidity $\l_{i}$ \\ \hline         
        Adjoint hypermultiplet mass $m$ & Interaction strength $m$ \\ \hline 
        Fundamental hypermultiplet mass $m_{k}$ &  Inhomogeneity and spin $p_{k} - s_{k}\e$  \\ \hline     
        Anti-fundamental hypermultiplet mass $\tilde m_{k}$ &  Inhomogeneity and spin $p_{k} + s_{k}\e$  \\ \hline    
        $\Omega$-deformation parameter $\e$ & Planck's constant $-i\hbar$ 
    \end{tabular}
  \end{center}
    \caption{Dictionary between gauge theories and integrable systems.}
    \label{T1}
\end{table}
In the next section, we examine in more detail how to obtain the integrable systems from the gauge theory data. 
\subsection{Classical integrable systems from compactified gauge theories}\label{CxIntSys}
The Coulomb branch of the elliptic quiver gauge theory is effectively an abelian theory with gauge group $U(1)^{r}$, where $r = KN - N+1$. Its dynamics is governed by the pre-potential $\mathcal{F}$, which can be determined by the Seiberg-Witten curve $\Sigma$ and the meromorphic one-form $\lambda_\text{~\!SW}$. The Coulomb branch moduli space is parametrized by periods of the Seiberg-Witten differential around homology cycles $\{A_I, B_I\}$ for $I = 1, \ldots, r$,
\eq{\vec a = \frac{1}{2\pi i}\oint_{\vec A} \lambda_\text{~\!SW}, \qquad \vec a^D = \frac{1}{2\pi i}\oint_{\vec B} \lambda_\text{~\!SW}.}
The low-energy effective theory takes the form of a non-linear sigma model on the Coulomb branch whose metric can be read off from the period matrix 
\eq{\tau_{IJ} = \frac{\partial \mathcal{F}}{\partial a^I \partial a^{J}}.}
This gives the Coulomb branch the structure of the base of a complex integrable system. The full structure of the Seiberg-Witten integrable system can be seen more clearly when we compactify the 4d theory on a circle down to three dimensions \cite{Seiberg:1996nz}. One can turn on Wilson loops around the compactified circle, which define real periodic scalars $\theta^I_e = {\oint A^I_3~\! dx^3}$. The dual photons define another set of real periodic scalars $\theta^I_m$. The 3d Coulomb branch is parametrized by the complex scalars $a^I$ as well as $\theta^I_e$ and $\theta^I_m$. It takes the form of a Jacobian fibration over the base manifold where the Jacobian torus is parametrized by $z_I = \theta_{m,~\!\!I} - \tau_{IJ}~\! \theta^J_e$. The variables $(a^I, z_I)$ play the role of the action-angle variables of the complex integrable system. The dynamics are linear flows on the Jacobian torus.

As suggested by Kapustin, the Coulomb branch of the compactified quiver gauge theory has an alternative description in terms of the Higgs branch of the mirror theory \cite{Kapustin:1998xn}. The Higgs branch does not receive quantum corrections and provides an alternative, simpler description of the theory. Let us review how to obtain the Higgs branch description of the theory via a sequence of string dualities. The branes are compactified in the $x^{3}-x^{6}$ torus. Performing a T-duality along the $x^{3}$ circle, the D4-branes become D3-branes along the $0126$ direction. S-duality interchanges the NS5-branes with D5-branes along the $012345$ direction. A further T-duality along the $x^{3}$ circle maps the D3-D5 system into a D4-D4' system with $K$ D4-branes wrapped around the $x^{3}-x^{6}$ torus and $N$ D4'-branes localized on the torus. This is the impurity theory \cite{Kapustin:1998pb}.

Alternatively, we can T-dualize along the $x^{6}$ circle to obtain a D2-D6 system where the $K$ D2-branes can be interpreted as $SU(N)$ instantons moving in the four transverse directions inside the D6-brane. Because the $x^{3}$ and the $x^{6}$ directions are compactified, the instantons live on $\mathbb{R}^2 \times T^2$. We further turn on Wilson lines for the $SU(N)$ gauge field on $T^2$, which specifies the positions of the impurities on the dual torus. The instanton moduli space describes the Higgs branch of the $SU(K)$ gauge theory, which in turn coincides with the Coulomb branch of the original gauge theory via the mirror map \cite{Intriligator:1996ex}.\footnote{The quantum mechanics on the moduli space of instantons on $\mathbb{R}^2 \times T^2$ has recently been studied in \cite{Dorey:2014kla}.}

This moduli space of $K$ $SU(N)$ instantons on $\mathbb{R}^2 \times T^2$ can then be mapped to a 2d theory living on the dual torus $\hat T^2$ using the Nahm transform. The 2d theory is described by a complex $U(K)$ gauge field $A_{z}(z, \bar z), A_{\bar z}(z, \bar z)$ and an adjoint scalar $\phi(z, \bar z)$. The fields have prescribed boundary conditions parametrized by the impurities $Q^\alpha_k, \tilde Q^\alpha_k$ in the fundamental and the anti-fundamental representations of $U(K)$ at the punctures $z_\a$. The infinite-dimensional space spanned by these fields is endowed with a canonical symplectic form
\eq{\omega_\infty= \int d^2z \sum^K_{k,\ell=1} d\phi_{k\ell}(z,\bar z) \wedge d A^z_{\ell k}(z, \bar z) +  \sum_{k=1}^K \sum_{\alpha = 1}^N d Q_k^\a \wedge d \tilde Q_k^\a. \label{symplectic}}
The F-term equations modulo the complexified gauge symmetry $U(K, \mathbb{C})$ give rise to the Hitchin equations
\eq{
\partial_{\bar z} \phi_{k\ell} + [A_{\bar z}, \phi]_{k\ell}= 2\pi i \sum^N_{\a=1}S^\a_{k\ell}\,\d^{(2)}(z-z_\a),
}
where the $GL(K, \mathbb{C})$ spin variables are defined as 
\eq{S^\a_{k\ell} = Q^\a_k \tilde Q^\a_\ell - \frac{m}{N}~\!\delta_{k\ell}.
\label{GL(K)variables}}
They are  subject to the constraints
\eq{\frac{1}{K}\sum_{k=1}^K  Q^{\a}_{k} \tilde Q^{\a}_{k} = m_\a, \qquad \sum_{\a=1}^N  Q^{\a}_{k} \tilde Q^{\a}_{k} = m.}
The first constraint implies that the spins lie in a particular conjugacy class of $GL(K, \mathbb{C})$ and the second constraint ensures that the diagonal components of $\phi_{k\ell}$ have vanishing residues.
This defines the hyper-K\"ahler quotient description of the instanton moduli space $\mathcal{M}_{K, N}$. 

Let us proceed to solve the Hitchin system defined on the dual torus.
With a suitable choice of gauge, one can diagonalize the connection as
\eq{A_{\bar z} = \frac{\pi i}{2(\bar\omega_2~\! \omega_1 - \bar \omega_1~\! \omega_2)} ~\!\text{diag}(x_1, x_2, \ldots, x_K),}
where $\omega_i$ are the half-periods of the dual torus and $x_k$ transform under a large gauge transformation into
\eq{
x_k \simeq x_k + 2n\, \omega_1 +  2m\, \omega_2, \qquad n,m\in \mathbb{Z}.
}
One can then decompose the Higgs field $\phi$ into diagonal and off-diagonal pieces and look for doubly-periodic solutions with simple poles at the impurities. The answer is given in terms of a combination of Weierstra$\ss$ functions as \cite{Nekrasov:1995nq, Dorey:2001qj}
\EQ{\phi_{k\ell}(z) &= \d_{k\ell}\left[p_{k} + \sum_{\a=1}^{N} S^{\a}_{kk}\,\z(z-z_{\a})\right]  +(1-\d_{k\ell})\sum^{N}_{\a=1} S^{\a}_{k\ell}\, \frac{\s(x_{k\ell}+z-z_\a)}{\s(x_{k\ell})\s(z-z_\a)}~\!e^{x_{k\ell}(\psi(z)-\psi(z_\a))}\label{gsCMLax},}
where $p_k$ are constants and we abbreviate $x_k  - x_\ell$ as $x_{k \ell}$. The function $\psi(z)$ is defined as 
\eq{\psi(z) = \frac{1}{\bar\omega_2~\! \omega_1 - \bar \omega_1~\! \omega_2}\big[\z(\omega_{2})(\bar \omega_{1}z - \omega_{1}\bar z) - \z(\omega_{1})(\bar \omega_{2}z - \omega_{2}\bar z)\big].}
One may readily check that this solution satisfies the required properties by recalling that the Weierstra$\ss$ functions are quasi-periodic functions that transform as
\eq{
\zeta(z+2~\!\omega_i) = \zeta(z) + 2~\!\zeta(\omega_i), \qquad \s(z+2~\!\omega_i) = -\s(z)~\! e^{2(z+\omega_i)~\!\z(\omega_i)},
}
and that $\z(z)$ has a simple pole at the origin while $\s(z)$ is regular there.

The canonical symplectic form (\ref{symplectic}) on the ambient space descends to the quotient space
\eq{\omega = \sum_{k=1}^K dx_k \wedge dp_k + \sum_{k=1}^K \sum_{\alpha = 1}^N dQ_k^\a \wedge d\tilde Q_k^\a.}
The symplectic form defines the Poisson bracket among the variables $x_k, p_k, Q^\a_k, \tilde Q^\b_k$
\eq{
\{x_{k}, p_{\ell}\} = \delta_{k\ell}, \qquad \{Q^{\a}_{k}, \tilde Q^{\b}_{\ell}\} = \d^{\a\b}\d_{k\ell}.
} 
The hyper-K\"ahler quotient implies that gauge-invariant quantities that trivially Poisson-commute in the ambient space will automatically commute in the quotient space. The trace of powers of $\phi$ will form an infinite tower of Poisson-commuting Hamiltonians
\eq{H_{n} = \frac{1}{n}~\!\text{tr}~\!\phi^{n}.} 
The Hamiltonians are encoded in the spectral curve, which is defined as the characteristic equation for $\phi$
\eq{\G(z,v) \equiv \det (v\, \mathbb{I} - \phi(z)) = 0.\label{spectralcurve}}
The meromorphic differential $v\,dz$ is the Seiberg-Witten differential. 
The generators $\phi$ of the commuting conserved charges are known as the Lax matrix in the integrable systems literature. The quotient construction gives the resulting manifold the structure of an algebraic integrable system \cite{Nekrasov:1995nq}. 

\subsection{The inhomogeneous spin Calogero-Moser model and its degenerate limits}

The classical integrable system describes $K$ non-relativistic particles on a doubly-periodic lattice which interact with pairwise elliptic potentials and spin exchanges. The Hamiltonian which contains the momentum quadratically as the non-relativistic kinetic term is given explicitly as \cite{Nekrasov:1995nq, Dorey:2001qj}
\EQ{&H = \sum^K_{k=1} p^2_k + \sum^K_{\ell \ne k} \sum^N_{\a=1}  S^\a_{k\ell} \, S^\a_{\ell k}\,\wp(x_{k\ell}) + \sum^K_{k=1}\sum^{N}_{\b \ne \a} S^\a_{k k} \, S^\b_{k k}\left(\wp(z_{\a\b}) - \z(z_{\a\b})^2\right)\\
& + \sum^K_{\ell \ne k} \sum^N_{\a \ne \b} S^\a_{k\ell} \, S^\b_{\ell k}\,\frac{\s(x_{k\ell}+z-z_\a)}{\s(x_{k\ell})\s(z-z_\a)}e^{x_{k\ell}z_{\a\b}}\Big[\z(x_{k\ell}+z_{\a\b}) - \z(x_{k\ell}) \Big] 
.}
The variables $x_k$ and $p_k$ correspond to the position and momentum of the Calogero-Moser particles. The $GL(K, \mathbb{C})$ spin variables can be traded with the $SL(N, \mathbb{C})$ spin variables $S_{k}^{~\!\!\a\b}$ labeling the internal spin carried by each particle
\eq{S^{\,\alpha\beta}_{k} = Q^{\a}_{k}\tilde Q^{\b}_{k} - \frac{\d^{\a\b}}{N}\sum_{\g=1}^N Q^{\g}_{k}\tilde Q^{\g}_{k}.\label{SL(N)variables}}
We will call the corresponding integrable model the inhomogeneous spin Calogero-Moser model. The two special gauge theories considered above correspond to special limits of this general model. In the following we will explore these models in detail.

There is an important subtlety that we need to address when quantizing the models. The integrable models that arise from gauge theory are \emph{complex} integrable systems where the dynamical variables $x_k, p_k$ are complex-valued and the spin variables $S^{~\!\!\a\b}_k$ take value in representations of the complex Lie algebra $\mathfrak{\mathfrak{sl}}(N, \mathbb{C})$. One can impose different reality conditions to obtain inequivalent real integrable systems. This chooses a middle-dimensional subspace of the complex integrable system. Upon quantization these lead to quantum integrable systems with very different wave functions and spectra. For example, the spins can be chosen to lie in a compact $\mathfrak{su}(N, \mathbb{R})$ representation or in a non-compact $\mathfrak{sl}(N, \mathbb{R})$ representation. One can also choose the particles to lie on the real axis or on the imaginary axis. In the asymptotic limit, we obtain the hyperbolic Sutherland or the trigonometric Sutherland model, respectively. The former has a continuous spectrum whereas the latter has a discrete spectrum. Because we wish to solve for the two-body S-matrices, we will choose the particles to lie on the real axis. We can then make a choice for the representation of the spin variables. As we will discuss in section \ref{section4}, this corresponds to conditions on the Seiberg-Witten curve where periods around various cycles become quantized. 

\paragraph{The spin Calogero-Moser model}
\

The model simplifies when we take all the inhomogeneities to coincide at the origin. In the brane picture, this corresponds to taking all the $N$ NS5-branes to be coincident. We reduce to the spin Calogero-Moser system \cite{Gibbons1984generalisation, Minahan:1992ht, Ha:1992zz, krichever1995spin}.
The Hamiltonian is given by
\eq{H = \sum_{k=1}^{K}\frac{p^{2}_{k}}{2} + \sum^K_{\ell > k}\sum_{\a,\b=1}^N S^{\a}_{k\ell}\, S^{\b}_{\ell k}\,\wp(x_k-x_\ell).
\label{sCM}
}
\paragraph{The Hubbard-Toda model}

\

The Calogero-Moser particles interact with a pairwise potential. There is a special limit where only the nearest neighbors interact while others are suppressed exponentially. This is obtained by taking the particles to be far apart while tuning the coupling such that 
\eq{x_{k} \to X_{k} + k \log \m^2, \qquad \Lambda^{K} = \m^{K}e^{2\pi i \t} \text{ fixed}.}
This is known as the Inozemtsev limit \cite{Inozemtsev1989finite}. In this limit, the Calogero-Moser system degenerates into the Toda chain where the particles interact via the nearest-neighbor exponential potential. The Inozemtsev limit of the inhomogeneous spin Calogero-Moser system gives rise to a new integrable system. The new system is a hybrid between the Toda chain and the Hubbard model.
The Lax matrix is given by (see appendix \ref{appendixA} for details)  
\eq{L_{k\ell}(z) = L^\text{Toda}_{k\ell}(z) + A_{k\ell}\left[\d_{k\ell} \frac{t+1}{2(t-1)} + \Theta_{k\ell}\frac{t}{t-1} + \Theta_{\ell k}\frac{1}{t-1}\right], \qquad t = e^{z}. \label{LaxHT}
}
Here $L^\text{Toda}$ is the Lax matrix for the Toda chain (\ref{TodaLax}), the hopping operators $A_{k\ell}$  (\ref{hopping}) satisfy 
\eq{
\{A_{k\ell}, A_{mn}\} = \delta_{kn}A_{m\ell}-\delta_{m\ell}A_{kn},
}
and $\Theta_{k\ell}$ is the discrete Heaviside function taking value $1$ for $k > \ell$ and zero otherwise.
The quadratic Hamiltonian can be written as
\EQ{H_2 &=H^\text{Toda} + \sum^{K}_{k=1}\frac{A_{kk}^{2}}{8} + \frac{t}{2(t-1)^{2}}\text{tr}~\! A^{2} + \frac{t+1}{2(t-1)} \sum^{K}_{k=1}p_{k}A_{kk} \\
&+ \sum^{K-1}_{k=1}e^{\frac{X_{k}-X_{k+1}}{2}}\left(\frac{t}{t-1} A_{(k+1)k} - \frac{1}{t-1} A_{k(k+1)} \right) + \Lambda^{K}e^{\frac{X_{K}-X_{1}}{2}}\left(\frac{t}{t-1}A_{1K} - \frac{1}{t-1}A_{K1} \right).
}
Here $H^\text{Toda}$ is the Toda Hamiltonian (\ref{TodaHamiltonian}). We define the Hubbard-Toda Hamiltonian such that the momentum only appears quadratically in the non-relativistic kinetic term. If we write $H_2$ as 
\eq{
H_2= \frac{\mathcal{H}_{2}t^{2} + \mathcal{H}_{1}t + \mathcal{H}_{0}}{(t-1)^{2}},
\label{HTquadratic}
}
then the coupled $p_{k}A_{kk}$ term can be eliminated by taking the linear combination $H^\text{HT} = (\mathcal{H}_2 + \mathcal{H}_0)/2$: 
\eq{
H^\text{HT}=H^\text{Toda} + \sum^{K}_{k=1}\frac{A_{kk}^{2}}{4} + \frac{1}{2}\Bigg[\sum^{K-1}_{k=1}e^{\frac{X_{k}-X_{k+1}}{2}}\left(A_{(k+1)k} + A_{k(k+1)} \right) + \Lambda^{K}e^{\frac{X_{K}-X_{1}}{2}}\left(A_{1K} + A_{K1} \right)\Bigg].\label{Hubbard-Toda}} 
When the particles are frozen to their equilibrium configuration, it is the Hubbard model describing a lattice with spins hopping between neighboring sites. Having obtained the Hamiltonian for (1) the spin Calogero-Sutherland model, and (2) the Hubbard-Toda model, our next goal is to diagonalize them. Traditionally, this can be achieved using the quantum inverse scattering method \cite{Faddeev:1979gh, korepin1997quantum}. We will review this method and apply it to the integrable models in the next section.
\section{Solution by the quantum inverse scattering method}\label{section3}
\begin{figure}[h]
\centering
\includegraphics[scale=0.3]{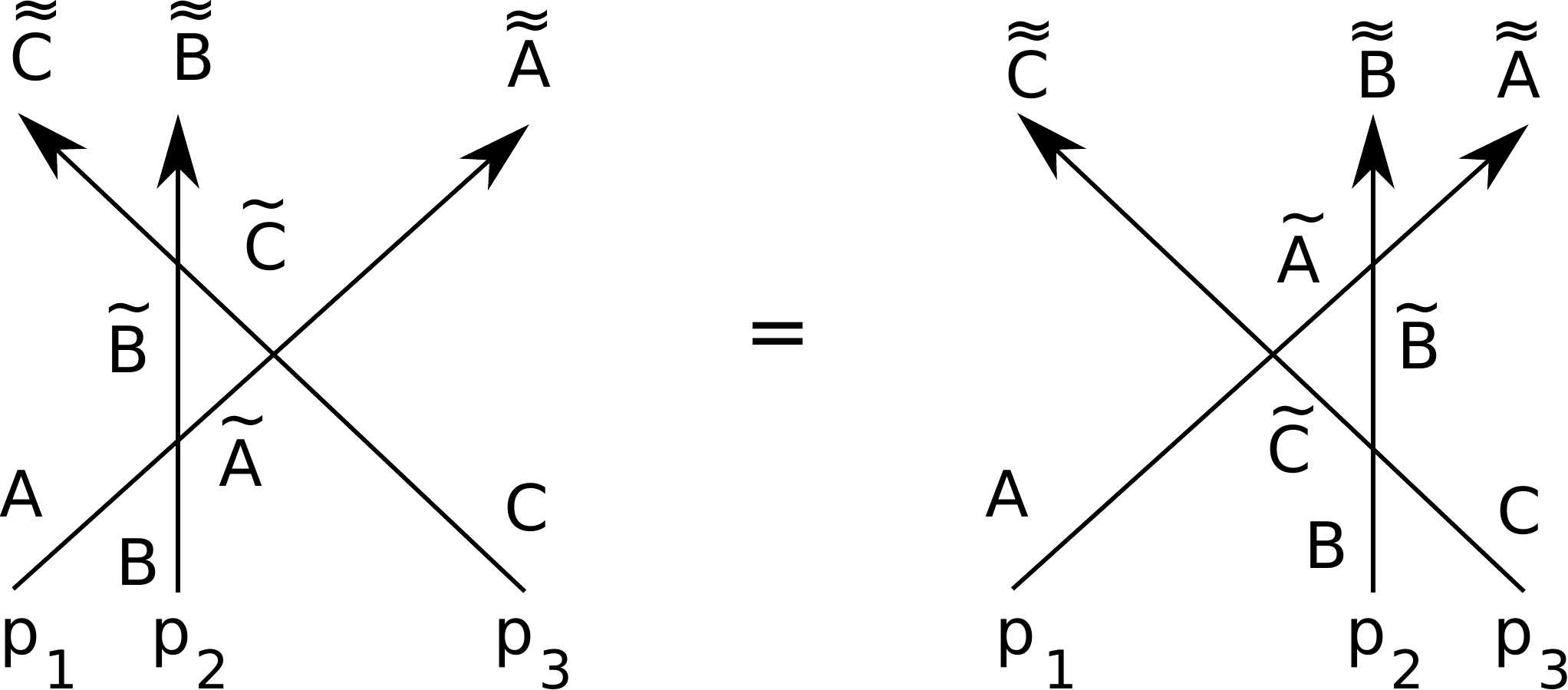}
\caption{The Yang-Baxter relation.}
\label{Yang-Baxter}
\end{figure}

The hallmark of quantum integrable models is factorized scattering,  i.e., particle scatterings factorize into a series of two-body scatterings and individual momentum is conserved. As the models we consider are $\mathfrak{sl}(N)$-invariant systems, each particle transforms in an irreducible representation $h^A$ of $\mathfrak{sl}(N)$ labeled by an index $A$. 
Due to translational invariance, the two-body S-matrix\, $\mathbb{S}: h^A \otimes h^B \to h^{\tilde A} \otimes h^{\tilde B}$ only depends on the difference of the momenta and takes the form
$\mathbb{S}^{\tilde A \tilde B}_{A B}(p_k-p_\ell)$. Consistency of scattering implies that the outcome is independent of the order in which the particles are scattered. This is shown in figure \ref{Yang-Baxter},
from which one may read off the Yang-Baxter equation 
\eq{\mathbb{S}^{\tilde{\tilde B} \tilde{\tilde C}}_{\tilde B \tilde C}(p_2-p_3)\,\mathbb{S}^{\tilde{\tilde A} \tilde C}_{\tilde A C}(p_1-p_3)\,\mathbb{S}^{\tilde A\tilde B}_{AB}(p_1-p_2) = \mathbb{S}^{\tilde{\tilde A}\tilde{\tilde B}}_{\tilde A \tilde B} (p_1-p_2)\,\mathbb{S}^{\tilde A \tilde{\tilde C}}_{A\tilde C}(p_1-p_3)\,\mathbb{S}^{\tilde B \tilde C}_{BC}(p_2-p_3)\label{qYB}.}
The Yang-Baxter equation strongly constrains the form of the S-matrix. It completely fixes the S-matrix up to a scalar factor: 
\eq{\mathbb{S}^{\tilde A \tilde B}_{A  B}(p) = S_0(p)\,\mathbb{R}^{\tilde A \tilde B}_{A B}(p).}
For example, when the particles are in the fundamental representation of $\mathfrak{su}(N)$ labeled by $\a = 1, \ldots, N$, the only invariant tensors are the identity operator $\mathbb{I}~\!^{\tilde \a\tilde \b}_{\a \b} = \delta^{\tilde \a}_\a\delta^{\tilde \b}_\b$ and the permutation operator $\mathbb{P}~\!\!^{\tilde\a\tilde\b}_{\a \b} = \delta_\a^{\tilde \b}\delta_\b^{\tilde \a}$. The R-matrix can be constructed from a linear combination of the invariant tensors $\mathbb{R} (p) = a(p)~\! \mathbb{I} + b(p)~\! \mathbb{P}$. The relative coefficients can then be constrained by the Yang-Baxter equation, and we obtain the fundamental $\mathfrak{su}(N)$ R-matrix
\eq{\mathbb{R}(p) = \frac{p~\!\mathbb{I} + i~\!\mathbb{P}}{p+i}.}

The scalar factor $S_0(p)$ does not depend on the Yang-Baxter equation and is more tricky to determine. It is usually fixed by imposing unitarity and crossing symmetry. Even then, there can be multiple solutions to the crossing equation and one needs to further impose conditions on the bound-state spectrum, e.g., the assumption of absence of poles and zeros in the physical momentum region. For particle models where the explicit Hamiltonians are known, we can in principle determine the full S-matrix by directly solving the Schr\"odinger equations. The components of the S-matrix can then be read off from the asymptotics of the wave functions. In practice, however, diagonalizing the matrix Schr\"odinger equations can be difficult. We will illustrate the use of both methods for the spin Calogero-Moser model and the Hubbard-Toda model in this section.

We begin by solving the problem on an infinite line. The asymptotic states are well-defined in the infinite-volume limit. Assuming that the particles are well-separated, the $K$-particle wave function can be written as
\eq{
\psi_{A_1\cdots A_K}(x_1, \ldots, x_K) \simeq \sum_{\s \in \text{Sym}(K)} \mathbb{S}_\s \,\exp\left(i\sum^K_{k=1} p_{\s(k)}\, x_k\right).
}
Here $\mathbb{S}_\s$ denotes that if two permutations $\s$ and $\s'$ differs by swapping $k$ and $\ell$, then $\mathbb{S}_\s/\mathbb{S}_{\s'} = \mathbb{S}_{k\ell}(p_k-p_\ell)$.
We then place the system on a large circle of radius $L \gg 1$ and impose a twisted periodic boundary condition
\eq{
\psi_{A_1\cdots A_K}(x_2, \ldots, x_K, x_1 + L) = \mathbb{T}^{-1}_{A_1 B_1} \psi_{B_1A_2\cdots A_K} (x_1, \ldots, x_K)
,}
where $\mathbb{T} = \exp(2\pi i\t S^z)$ is a twisted periodic boundary condition that rotates the spin at the end of the chain in the preferred $z$-direction. For $L \gg 1$, we have 
\eq{\psi_{A_1\cdots A_K}(x_2, \ldots, x_K, x_1 + L) \simeq e^{-ip_1 L}\, \mathbb{S}_{~\!\!12}(p_1-p_2)\cdots \mathbb{S}_{~\!\!1K}(p_1 - p_K)~\! \psi_{A_1\cdots A_K}(x_1, \ldots, x_K).} 

The asymptotic Bethe ansatz dictates that the phase shift acquired by a particle around the system is equal to the product of S-matrices with all the other particles. 
\eq{e^{ip_{k}L}|\psi\rangle =\mathbb{T} \prod^{K}_{\ell\ne k}\mathbb{S}_{~\!\!k\ell}(p_{k}-p_{\ell})|\psi\rangle.\label{operator}}
The solution in terms of the asymptotic Bethe ansatz is valid for large system size $L \gg 1$ up to sub-leading finite-size effects of order $\exp(-8\pi^2/g^2)$. This is also the starting point of the finite-volume problem, where we go to the mirror model and formulate a set of thermodynamic Bethe ansatz equations. As we shall explain below, this is closely related to the diagonalization problem of the integrable spin chain.

\subsection{The spin Calogero-Moser model}\label{sCSsolution}
Let us solve the spin Calogero-Moser model (\ref{sCM}) using the asymptotic Bethe ansatz. This model is quantum integrable \cite{hikami1993integrability}. In the large-volume limit, one period in the lattice grows to infinity. The Hamiltonian becomes that of the spin Calogero-Sutherland model.
\eq{H = \sum_{k=1}^{K}\frac{p^{2}_{k}}{2} + \sum^K_{\ell > k}\sum_{\a,\b=1}^N\frac{S^{\a}_{k\ell}\, S^{\b}_{\ell k}}{4\sinh^{2}\frac{x_{k\ell}}{2}},
\label{sCS}
}
The particles now move on the real line and interact with a hyperbolic potential. The Hamiltonian can be rewritten in terms of the $\mathfrak{sl}(N)$ spin variables using (\ref{GL(K)variables}) and (\ref{SL(N)variables}) as
\eq{
\sum_{\a,\b=1}^N S^{\a}_{k\ell}\, S^{\b}_{\ell k} 
=\sum_{\a,\b=1}^N S^{\a\b}_{k} S^{\b\a}_\ell + \frac{m^2}{N}.
}
Note that this coincides with the square of the total spin when $N=2$:
\eq{
\left(S_{k} + S_{\ell}\right)^2 
=\sum_{\a,\b=1}^N S^{\a\b}_k S^{\b\a}_{\ell} + m^2-\frac{m^2}{N}
.}

To quantize the model, we promote the spin variables $S_k$ to the spin operators $\hat S_k$.
For simplicity, we take all sites to be in the same spin-$s$ representation of $\mathfrak{su}(2)$. Integrability implies that the particle interactions of the model can be factorized into a series of two-body scatterings. It is useful to go to the center-of-mass frame such that
$p_{1} = - p_{2} = p/2$ and $x = x_{2} - x_{1}$.
The S-matrix can be determined by solving the matrix Schr\"odinger equation
\eq{\left(-\frac{d^{\, 2}}{dx^2} \mathbb{I} + \mathbb{V}(x)\right) \vec\psi(x) = E~\!\vec\psi(x), \label{matrixSchrodinger}}
where the potential is written in terms of the total spin operator as 
\eq{\mathbb{V}(x) = \frac{\left(\hat S_1 + \hat S_2 \right)^2}{4\sinh^{2}\left(\frac{x}{2}\right)}.}
In each irreducible subspace of the total spin operator labeled by the eigenvalue  $s$, we obtain a scalar Schr\"odinger problem with a hyperbolic potential and coupling $s(s+1)$. The scattering phase of particles in a hyperbolic potential is known explicitly \cite{Olshanetsky:1981dk} 
\eq{\mathbb{S}(p) = \frac{\G(1+\hat S-i p)~\!\G(1+ i p)}{\G(1+\hat S+ip)~\!\G(1-i p)}\, \mathbb{P}.\label{Smatrix}}
The key observation is that the Calogero-Sutherland particle S-matrix (\ref{Smatrix}) coincides with the universal R-matrix of the Heisenberg $\mathfrak{su}(2)$ spin chain \cite{Kulish:1981gi}. In this case the scalar part of the S-matrix is  
\eq{S_0(p) = \frac{\G(1+2s-i p)~\!\G(1+ i p)}{\G(1+2s+ip)~\!\G(1-i p)},}
such that $\mathbb{R}(p)$ acting on the ground state with all spins up is the identity operator. We can then proceed to solve the model for a set of Bethe ansatz equations using the quantum inverse scattering method.

Consider a ghost particle labeled by 0 with momentum $p$ scattering with the other $K$ particles. We define the fundamental monodromy matrix $\hat T_0(p)$ as the product of R-matrices and the transfer matrix $\hat t(p) = \text{tr}_{0}~\! \hat T_0(p)$ as its trace over the ghost particle:
\eq{\hat T_0(p) =  \mathbb{T}\prod^{K}_{k=1} \mathbb{R}_{~\!\!0k}(p-p_{k}).\label{monodromy}}
Integrability of the model follows because the fundamental monodromy matrix likewise satisfies the Yang-Baxter equation. This then implies that the fundamental transfer matrices commute at different values of the spectral parameters $[\hat t(p), \hat t(p')] = 0$, and generate the quantum commuting Hamiltonians.

\begin{figure}[h]
\centering
\includegraphics[scale=0.3]{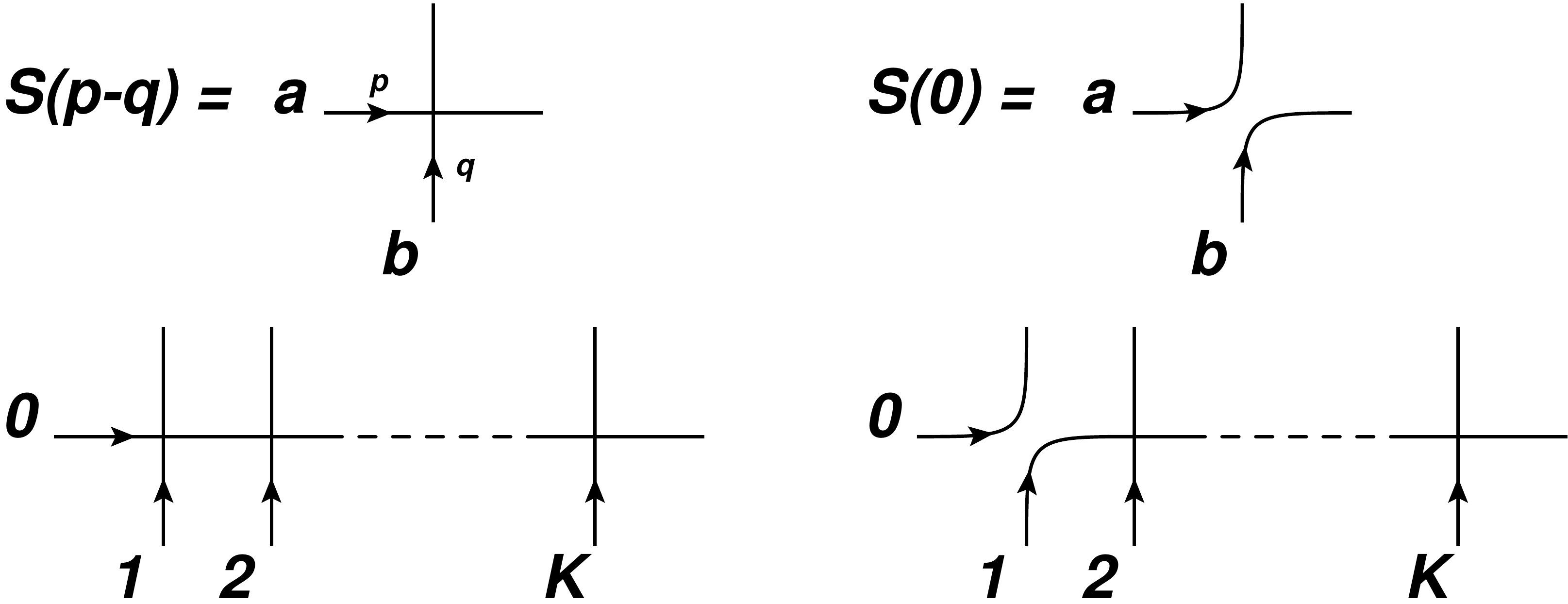}
\caption{The S-matrix becomes the permutation operator when the two particles have the same momentum. The transfer matrix describes the scattering of a ghost particle with the physical particles. When the ghost particle has the same momentum as a physical particle, they can be swapped and the fundamental transfer matrix becomes the scattering of that physical particle with the other physical particles.}
\label{transferfigure}
\end{figure}

If we evaluate the fundamental monodromy matrix at one of the particle momenta, then because $\mathbb{S}_{~\!\!0\ell}(0) = \mathbb{P}_{0\ell}$, the ghost particle swaps with that physical particle. For $p=p_1$, we may commute $\mathbb{P}_{01}$ to the right using  $\mathbb{P}_{0\ell}\,\mathbb{P}_{0k}\,\mathbb{P}_{0\ell} =  \mathbb{P}_{k\ell}$ to obtain
\eq{\prod^{K}_{k=1}\mathbb{S}_{~\!\!0k}(p_{1}-p_{k}) 
=\mathbb{S}_{~\!\!12}(p_{1}-p_{2})\,\mathbb{S}_{~\!\!13}(p_{1}-p_{3})\cdots\,\mathbb{S}_{~\!\!1K}(p_{1}-p_{K})\,\mathbb{P}_{01},
}
and similarly for arbitrary $p=p_{k}$.
We recover the products of S-matrices in (\ref{operator}) by tracing out the ghost particle since $\text{tr}_{0}~\!\mathbb{P}_{01} = 1$. Graphically, this can be represented as in figure \ref{transferfigure}.

Thus diagonalizing the product of S-matrices is equivalent to diagonalizing the fundamental transfer matrix. 
The latter problem can be solved using the nested algebraic Bethe ansatz \cite{Faddeev:1994nk, Faddeev:1996iy}. In this approach, we first introduce an auxiliary $\mathfrak{su}(2)$ space for the ghost particle. We define the auxiliary monodromy matrix $\hat T_a$ as in (\ref{monodromy}), where now the ghost particle lives in the auxiliary space. When the physical particles are in the spin-$1/2$ representation, the fundamental monodromy matrix coincides with the auxiliary monodromy matrix. The fundamental and the auxiliary monodromy matrices satisfy the Yang-Baxter equation
\eq{
\mathbb{R}_{a0}(p-q)\, \hat T_a(p)\, \hat T_0(q) = \hat T_0(q)\, \hat T_a(p)\, \mathbb{R}_{a0} (p-q),
}
implying that $[\hat t(p), \hat t_a(p')] = 0$ thus we can simultaneously diagonalize $\hat t$ and $\hat t_a$. It turns out to be simpler to first find eigenstates of the auxiliary transfer matrix and then calculate the eigenvalue of the fundamental transfer matrix on these eigenstates. We write the
auxiliary monodromy matrix as a $2 \times 2$ matrix over the auxiliary space where each entry is an operator acting on the physical space
\eq{\hat T_a(p) = \begin{pmatrix}\hat A(p) & \hat B(p) \\ \hat C(p) & e^{2\pi i\t}~\!\!\hat D(p) \end{pmatrix},}
such that the transfer matrix is simply $\hat t_a = \hat A + e^{2\pi i\t} \hat D$. The entries of the auxiliary monodromy matrix can be found using the explicit expressions of the R-matrix
\EQ{\mathbb{R}_{0k}(p) &= \frac{1}{p+i} \left(p\, \mathbb{I} + i \, \vec \sigma \cdot \vec S_k\right) \\
&= \frac{1}{p+i}\begin{pmatrix}
p + i S^z_k & i S^-_k\\
i S^+_k & p -i S^z_k &
\end{pmatrix}.
}
Thus we define the nested pseudo-vacuum $\Omega(\{p_k\})$ as the highest-weight state annihilated by $\hat C(p)$. It is also an eigenstate of the transfer matrix 
\eq{\hat A(p)\,\Omega = \prod_{k=1}^K \frac{p-p_k+is}{p-p_k+i}\,\Omega, \qquad \hat D(p)\,\Omega = \prod_{k=1}^K \frac{p-p_k-is}{p-p_k+i}\,\Omega.}
The eigenstates of $\hat t_a$ are generated by acting on the nested pseudo-vacuum with the creation operators labeled by the magnon rapidities $\l_i$:
\eq{\Phi(\{\lambda_i\}, \{p_k\})  = \hat B(\l_1)\cdots \hat B(\l_M)\,\Omega(\{p_k\}).}
The Yang-Baxter equation (\ref{qYB}) implies that the commutation relations are
\EQ{\hat A(p)\hat B(\l) &= \frac{p-\l-i}{p-\l} \hat B(\l)\hat A(p) + \frac{i}{p-\l}\hat B(p)\hat A(\l) \\
\qquad \hat D(p)\hat B(u) &= \frac{p-\l+i}{p-\l} \hat B(\l)\hat D(p) - \frac{i}{p-\l}\hat B(p)\hat D(\l).}
By commuting $\hat A$ and $\hat D$ past the creation operators using the commutation relations, we find that the eigenvalue of the auxiliary transfer matrix is 
\eq{t_a(p) = \prod^{K}_{k=1}\frac{p-p_k+is}{p-p_k+i}\prod^{M}_{i=1}\frac{p-\l_{i}-i}{p-\l_{i}} + e^{2\pi i \t}\prod^{K}_{k=1}\frac{p-p_k-is}{p-p_k+i}\prod^{M}_{i=1}\frac{p-\l_{i}+i}{p-\l_{i}}.\label{baxter}}
The eigenvalue of the fundamental transfer matrix on the Bethe state $\Phi(\{\lambda_i\}, \{p_k\}) $ is known to be \cite{Kulish:1981gi}
\eq{
t(p) = \prod^{M}_{i=1} \frac{p-\l_{i}-is}{p-\l_{i}+is} + \mathcal{O}(p^K).
}
We arrive at the asymptotic Bethe ansatz equation from (\ref{operator})
\eq{e^{ip_{k}L} =\prod^{K}_{\ell=1}\frac{\G(1+2s-i p_{k\ell})~\!\G(1+ ip_{k\ell})}{\G(1+2s+i p_{k\ell})~\!\G(1-ip_{k\ell})}\prod^{M}_{i=1} \frac{p_{k}-\l_{i}-is}{p_{k}-\l_{i}+is}.\label{BAEsCM1}}
We see that the eigenvalue of the S-matrix factorizes into two parts: one that represents the scattering of the particles and one that represents the scattering of the particles with the magnons. There is another set of Bethe ansatz equations for the magnons. This follows from requiring the auxiliary transfer function (\ref{baxter}) to have vanishing residue at $p = \l_{i}$.
\eq{\prod^{K}_{k=1} \frac{\l_{i}-p_{k} + is}{\l_{i}-p_{k} - is} = e^{2\pi i \t}\prod^{M}_{j \ne i} \frac{\l_{i}-\l_{j} + i}{\l_{i}-\l_{j} - i}.\label{BAEsCM2}}
This is the Bethe ansatz equation describing the scattering of magnons on a spin chain with inhomogeneity $p_k$ at each site. 

\subsection{The Hubbard-Toda model}
The Hubbard-Toda model is a dynamical lattice of Toda particles with spins that can hop between neighboring sites. This new feature also makes diagonalizing the problem difficult. The model is classically integrable as it arises from the Inozemtsev limit of the inhomogeneous spin Calogero-Moser system. We explicitly verify the classical integrability of this model in appendix \ref{appendixB}. While we do not yet have a proof of quantum integrability, we will assume that the model is also integrable at the quantum level and solve for the two-body S-matrix. 

\subsubsection{The non-compact $\mathfrak{sl}(2)$ model}

We begin by focusing on the non-compact model where each particle carries canonically-commuting bosonic oscillators $Q^{\alpha}_{k} = (a_{k}, b^{\dagger}_{k}), \, \tilde Q^{\alpha}_{k} = (a^\dagger_{k}, -b_{k})$ such that
\eq{[Q^{\alpha}_{k},\tilde Q^{\beta}_{\ell}] = \delta_{k \ell}\,\delta^{\a\b} \quad \Longleftrightarrow \quad [a_{k}, a^{\dagger}_{k}]=[b_{k}, b^{\dagger}_{k}]=1.}
The $\mathfrak{sl}(2, \mathbb{R}) \simeq \mathfrak{su}(1,1)$ algebra is generated by the spin variables $S^{~\!\!\a\b}_k$ (\ref{SL(N)variables}). 
In matrix form,
\eq{S^{~\!\!\a\b}_{k} = \begin{pmatrix} \frac{1}{2} \left(N^a_{k} + N^b_k+1\right) & -a_{k} b_{k}\\
b^{\dagger}_{k}a^{\dagger}_{k}  & -\frac{1}{2} \left(N^a_{k} + N^b_k+1\right)
\end{pmatrix},}
where $N^a_k = a^\dagger_k a_k$ and $N^b_k = b^\dagger_k b_k$ are the number operators. The quadratic Casimir is $S^2_k = s_k(s_k+1)$, where $s_{k} = (N^{b}_{k}-N^{a}_{k}-1)/2$.

For two sites, we take the total spin operator $S_G = S_1 + S_2$ as the global $\mathfrak{su}(1,1)_G$ generator. The quadratic Casimir $S^2_G = s_G(s_G+1)$ is determined by the tensor product 
\eq{V_{s_1} \otimes V_{s_2} = \bigoplus^\infty_{s_G=s_1+s_2+1} V_{s_G}.}
The two-body Hamiltonian is (\ref{Hubbard-Toda})
\eq{H^\text{HT} = \left(\frac{p}{2}\right)^{2}+\frac14(A^{2}_{11}+A^{2}_{22}) + \frac12 (A_{12} + A_{21})~\!e^{-x/2} + e^{-x}.
\label{HTHamiltonian}
} 
Our goal is to diagonalize the hopping term $A_{12} + A_{21}$. We write it as the sum of two terms $T_+ = A_{12}$ that moves a type-$a$ spin from the first site to the second site and moves a type-$b$ spin from the second site to the first site, and $T_- = A_{21}$, which does the opposite. First, observe that if we define the difference between spins at two sites as $T_z = s_{1} - s_{2}$, then $T_+$ increases $T_z$ by one unit and $T_-$ decreases $T_z$ by one unit. Hence $T_\pm$ and $T_z$ are generators of an auxiliary $\mathfrak{su}(2)$ symmetry algebra
\eq{[T_z, T_\pm] = \pm T_\pm, \qquad [T_+, T_-] = 2~\!T_z.} 
The total spin $\overline s = s_{1} + s_{2} + 1$ is conserved under the action by $T_\pm$ so commutes with them. The Hamiltonian (\ref{HTHamiltonian}) can then be written in terms of the auxiliary $\mathfrak{su}(2)$ generators as 
\eq{H^\text{HT} = \left(\frac{p}{2}\right)^{2}+\frac{1}{4}\left(s^{2}_A-T_z^2\right)+ T_x \,e^{-x/2} + e^{-x}\label{HTHamiltonian2}.}
The auxiliary $\mathfrak{su}(2)$ symmetry generators commutes with the global $\mathfrak{su}(1,1)$ symmetry generators. Remarkably, they  have the same quadratic Casimirs and are labeled by the same $s$.
This implies that the tensor product of two states can be decomposed as
\eq{\bigoplus^{\infty}_{s_{1}=-\frac{1}{2}}V_{s_{1}} \otimes \bigoplus^{\infty}_{s_{2}=-\frac{1}{2}}V_{s_{2}} = \bigoplus^{\infty}_{s=0} V_{s}\otimes W_{s},}
where $W_{s}$ is a spin-$s$ representation of the auxiliary $\mathfrak{su}(2)$ symmetry that counts the degeneracies of the spin-$s$ representation of the global $\mathfrak{su}(1,1)$ symmetry. For concreteness, we present the multiplet structure in table \ref{bases2}. 

\begin{table}[htbp]
\begin{center}
    \begin{tabular}{c c c}
        \hline
        $\mathfrak{su}(1,1)_{G}$ & $\mathfrak{su}(2)_{A}$ & Lowest-weight states \\ \hline\hline \\
        0 & $0$ & $(0,0)$  \\ \\
        1 & $1$ & $(B,A)$ \\ \\
        2 & $2$ & $(B^{2},A^{2})$  \\ \\
        \vdots & \vdots & \vdots
    \end{tabular}
  \end{center}
    \caption{Bases of two-particle states in the $\mathfrak{su}(1,1)$ Hubbard-Toda model. For ease of notation, we use $(A^{N^{a}_{1}}B^{N^{b}_{1}}, A^{N^{a}_{2}}B^{N^{b}_{2}})$ to denote a state with $N^{a}_{k}$ type-$a$ spins and $N^{b}_{k}$ type-$b$ spins at site $k$. The ladder operators $T_\pm$ hop spins between the two sites.}
    \label{bases2}
\end{table}
The S-matrix acting on the tensor product also decomposes into
\eq{
\mathbb{S}(p) = \sum_{s=\overline s}^{\infty} \mathcal{P}_{s}\, \mathbb{S}_{~\!\!s}(p),
}
where $\mathcal{P}_s$ projects onto the irreducible subspace $V_s$. $\mathbb{S}_s(p)$ is the operator acting on $W_s \otimes C^\infty(\mathbb{R})$, naturally given by the $(2s+1)$-component wave function $\vec{\psi} = (\psi_s, \ldots, \psi_{-s})$ and obeying the matrix Schr\"odinger equation (\ref{matrixSchrodinger}) with the Hubbard-Toda potential (\ref{HTHamiltonian2}).

Note that the linear combination $H_1 = -\mathcal{H}_2+\mathcal{H}_0$ from (\ref{HTquadratic}) defines another commuting Hamiltonian 
\eq{H_{1} = \frac{1}{2}\,T_z\, p - T_y\,e^{-x/2}, \label{constraint}}
that asymptotes to $T_z\, p/2$ as $x \to \infty$.
Because of the existence of the conserved charge $H_1$, $\Delta s = T_z$ is preserved in scattering. The S-matrix does not mix the various components of the wave function labeled by $\Delta s$ and can be written in a block diagonal form with $S(p; \Delta s, s)$ on the diagonals
\eq{\mathbb{S} (p) = \text{diag}(S(p; s, s), S(p; s-1, s), \ldots, S(p; -s, s)). \label{blockdiagonal}}
The form of the S-matrix can be fixed assuming integrability. The Yang-Baxter relation (\ref{qYB}) fixes the form of $\mathbb{S}_s(p)$ up to an overall scalar factor
\eq{
\mathbb{S}(p) = S_0(p)\sum_{s=\overline s}^{\infty} \mathcal{P}_{s}\, \mathbb{R}_{s}(p).
}
Here $\mathbb{R}_s(p)$ is given by the universal $\mathfrak{sl}(2)$ R-matrix \cite{Kulish:1981gi}
\eq{\mathbb{R}_s(p) = (-1)^{s-\overline s}~\!\frac{\G(-s-i p)~\!\G(-\overline s + i p)}{\G(-s+ip)~\!\G(-\overline s-i p)},}
where we have chosen the normalization such that $\mathbb{R}_{\overline s} = 1$. 
The model can then be solved using the quantum inverse scattering method by diagonalizing the transfer matrix, as discussed in the last section, leading to the following set of nested Bethe ansatz equations
\EQ{
e^{i p_k L} &= \prod^{K}_{\ell \ne k}S_{0}(p_{k}-p_{\ell}; s_{k},s_{\ell})\prod^{M}_{i=1}\frac{p_{k} - \l_{i} + is_{k}}{p_{k} - \l_{i} - is_{k}}  \\
\prod^{K}_{k=1}\frac{\l_{i} - p_{k} + is_{k}}{\l_{i} - p_{k} - is_{k}} &= e^{2\pi i\t}\prod^{M}_{j\ne i}\frac{\l_{i}-\l_{j} + i}{\l_{i}-\l_{j} - i}.
\label{HTBAE}
}
The Yang-Baxter equation alone, however, does not determine the scalar factor in the S-matrix. To fix the scalar factor, one needs to directly solve the matrix Schr\"odinger equation in the Hubbard-Toda potential, as we have done for the spin Calogero-Sutherland potential in the previous section. When we attempt to do so, we encounter a difficulty that the potential is not diagonal and the matrix becomes larger for higher spins. We will attempt to tackle the problem directly for the lower spins in the next section. 
As we shall see, an indirect method using the Bethe/Gauge correspondence gives a prediction for the full scalar factor for arbitrary spins. It is given by 
\eq{S_{0}(p; s_{k}, s_{\ell}) = \frac{\G\left(1+\Delta s+ip \right)\G\left(1-\Delta s+ip \right)\G\left(-\bar s - ip\right)}{\G\left(1+\Delta s-ip \right)\G\left(1-\Delta s-ip \right)\G\left(-\bar s + ip\right)}.}
Note that the $\overline s$-dependent parts in the scalar factor exactly cancel with those in the universal R-matrix, such that the S-matrix is a function only of $\Delta s$ and $s$.
\eq{\mathbb{S}(p; \Delta s, s)  = \frac{\G\left(1+\Delta s+ip \right)\G\left(1-\Delta s+ip \right)}{\G\left(1+\Delta s-ip \right)\G\left(1-\Delta s-ip \right)} \sum_{s=\overline s}^\infty \mathcal{P}_s ~\!\frac{\G\left(- s - ip\right)}{\G\left(- s +ip\right)}.
\label{prediction}}
The result agrees with the direct method and provides a convincing evidence that the Bethe/Gauge correspondence is an effective tool for solving quantum integrable systems. 

\subsubsection{Solving the matrix Schr\"odinger equation}
Let us attempt to directly diagonalize 
\eq{H^\text{HT} = \left(\frac{p}{2}\right)^{2}+\frac{1}{4}\left(s^{2}_A-T_z^2\right) + T_x\, e^{-x/2} + e^{-x}\label{h2},}
by solving for the exact wave function and reading off the scattering phase $S(p;\Delta s, s_A)$ from the asymptotics.
The first difficulty we encounter is that $T_z$ and $T_x$ cannot be simultaneously diagonalized in general, and we have to solve a matrix Schr\"odinger equation.  
Asymptotically, $H^\text{HT} \to  \left(p/2\right)^{2} + \left(s^{2}_A-T_z^2\right)/4$
so it acts as a free Hamiltonian on the spin $\D s$ component of the wave function $\vec\psi$. 
This determines the dispersion relation:
\eq{E = \left(\frac{p_{\Delta s}}{2}\right)^{2} + \frac{s^{2}_A-(\Delta s)^{2}}{4}.}
\paragraph{Spin 0:}For spin $0$, the problem reduces to a scalar Schr\"odinger equation with the Liouville potential
\eq{-\psi''(x)+e^{-x}\psi(x) = \left(\frac{p}{2}\right)^{2}\psi(x).}
The wave function is exactly solvable in terms of modified Bessel function of the second type. Up to an arbitrary normalization,
\eq{\psi(x) = K_{ip}(2e^{-\frac{x}{2})}.}
It oscillates at $x\to \infty$ and exponentially decays at $x \to -\infty$ as is expected from the Liouville potential.
The S-matrix can be read off from the ratio of the left-moving and the right-moving modes at $x \to \infty$
\eq{S(p;0,0) = -\frac{\Gamma \left(1+i p\right)}{\Gamma \left(1-i p\right)}. }
\paragraph{Spin 1/2:}
For the spin-$1/2$ representation, $T_x, T_y, T_z$ are the Pauli matrices and we obtain a system of two coupled Schr\"odinger equations. We can decouple the equations in the basis $\psi_{\pm} = \psi_{1/2} \pm \psi_{-1/2}$ in which $T_x$ is diagonal
\eq{-\psi_{\pm} ''(x)+\left(e^{-x} \pm \frac{1}{2}e^{-\frac{x}{2}}\right) \psi_{\pm} (x) = \left(\frac{p}{2}\right)^{2}\psi_{\pm}(x). 
}
It is an exactly solvable potential of Morse type and the wave functions are given in terms of confluent hypergeometric functions as
\eq{
\psi_{\pm}(x) = e^{-2 e^{-\frac{x}{2}}-\frac{ip~\!\!x}{2}}~U\!\left(\frac{1}{2}\pm \frac{1}{2}+ i p,1+2 i p,4 e^{-\frac{x}{2}}\right).}
Looking at its asymptotics as $x \to \infty$, 
we can read off the S-matrix as before
\eq{S\left(p;\pm\frac{1}{2},\frac{1}{2}\right) = \frac{\Gamma \left(\frac{1}{2}+ i p\right)}{\Gamma \left(\frac{1}{2}- i p\right)},}
which agrees precisely with our prediction (\ref{prediction}).
Note that the S-matrix has a simple pole at $p=i/2$, which is usually indicative of a bound state. However, $\psi_-$ with this value of $p$ grows instead of decays at infinity. One can see why this is so because the state has energy $-1/16$, which touches the bottom of the potential well. It has no zero-point energy so cannot be a bound state.
\paragraph{Spin 1:}
For the spin-1 representation, we have to solve the $3 \times 3$ coupled matrix Schr\"odinger equations \eq{\begin{pmatrix}-d^{~\!\!2}+ e^{-x} & \frac{1}{\sqrt2}e^{-\frac{x}{2}}& \\\frac{1}{\sqrt2}e^{-\frac{x}{2}} & -d^{~\!\!2} + e^{-x}+\frac{1}{4}& \frac{1}{\sqrt2}e^{-\frac{x}{2}}\\ & \frac{1}{\sqrt2}e^{-\frac{x}{2}}& -d^{~\!\!2}+e^{-x} \end{pmatrix}\begin{pmatrix}\psi_{1} \\ \psi_{0} \\ \psi_{-1}\end{pmatrix} = E \begin{pmatrix}\psi_{1} \\ \psi_{0} \\ \psi_{-1}\end{pmatrix}.\label{coupled}}
Note that if we define 
$\psi_{\pm} = \left(\psi_{1} \pm \psi_{-1}\right)/\sqrt{2}$, 
then the equation for $\psi_{-}$ decouples and 
the problem reduces to a $2 \times 2$ Hamiltonian acting on $\vec \psi=(\psi_{+}, \psi_{0})$,
\eq{H(x)\vec\psi(x) = E~\!\vec \psi(x), \qquad H(x) = \begin{pmatrix}-d^{~\!\!2} + e^{-x} & e^{-\frac{x}{2}} \\ e^{-\frac{x}{2}} & -d^{~\!\!2}+e^{-x}+\frac{1}{4} \end{pmatrix}.\label{coupled2}}

\begin{figure}
\begin{center}
\includegraphics{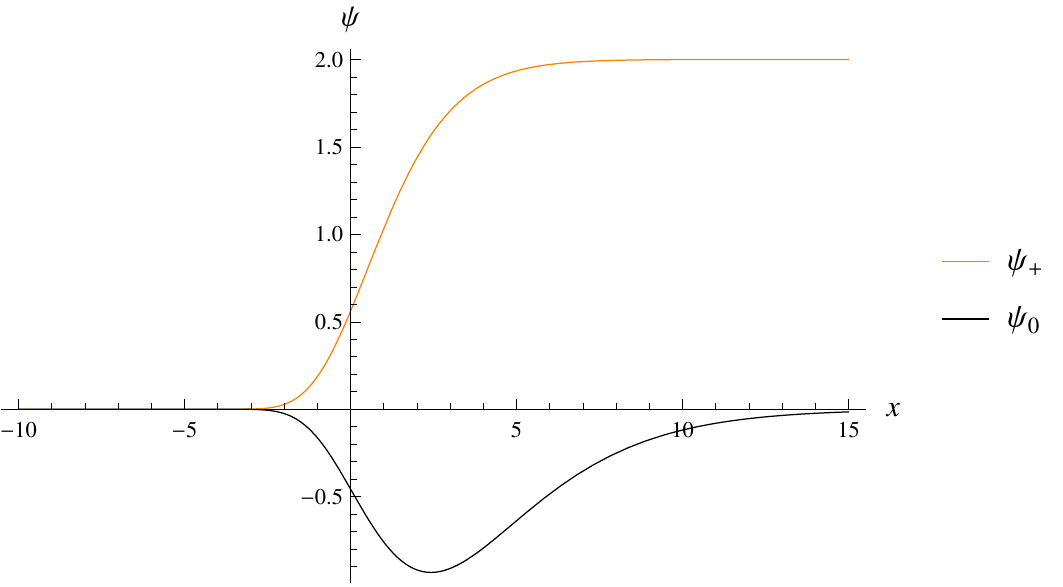}
\caption{The exact wave function for the spin-1 problem analytically continued to $p_{1} = 0$.}
\label{Plot1}
\end{center}
\end{figure}

We perform a Darboux (supersymmetric) transform to decouple the equations, and recover the eigenvector $\vec \psi$ from the diagonal basis. This goes as follows \cite{humi1985separation, Cannata:1993qj}. We look for a matrix $Q(x) = d+A(x)$ such that the new state 
$\vec\phi(x) = Q(x)~\!\vec\psi(x)$
is an eigenstate of a diagonalized Hamiltonian $\widetilde H$ with energy $E$.
In the language of supersymmetric quantum mechanics, $Q$ is the supersymmetric transform that intertwines the pair of supersymmetric Hamiltonians as $Q H = \widetilde H Q$.
Solving for $Q$ and $\widetilde H$, we find
\eq{
Q(x) = \begin{pmatrix}d & e^{-\frac{x}{2}}\\ e^{-\frac{x}{2}} & d+\frac{1}{2}\end{pmatrix}, \qquad \widetilde H(x) = \begin{pmatrix}-d^{~\!\!2}+e^{-x} & & \\ & -d^{~\!\!2}+e^{-x}+\frac{1}{4} \end{pmatrix}.
}
The decoupled equations reduce to the spin-0 problem and can be easily solved as before. The original wave functions can be obtained from $\vec \phi$ by inverting the Darboux transform:
\EQ{
\psi_{+}(x)&=y \big[2 c_{0}K_{i p_0}(y)- c_{+}\left(K_{i p_1-1}(y)+K_{i p_1+1}(y)\right)\big], \qquad y = 2~\!\! e^{-\frac{x}{2}}\\
\psi_{0}(x)&=y \left[2 c_{+}K_{i p_1}(y)+c_{0}\left(e^\frac{x}{2} K_{i p_0}(y)-K_{i p_0-1}(y)-K_{i p_0+1}(y)\right)\right].\label{wave function}}
$\psi_-$ can also be solved because it is decoupled from $\psi_+$ and $\psi_0$. Up to a normalization constant,
\eq{\psi_-(x) = c_- K_{ip_1}(2e^{-\frac{x}{2}}).}
The S-matrix can again be read off from the asymptotics of the wave functions. It agrees with our prediction when $c_- = 0$
\eq{S(p_{1}; \pm 1, 1) = \frac{\Gamma (1+i p_{1})}{\Gamma (1-i p_{1})},\qquad S(p_{0};0,1) = -\frac{(1-i p_{0})~\! \Gamma (1+ i p_{0})}{(1+ i p_{0})~\!  \Gamma (1- i p_{0})}.}
One may ask if there is any bound state corresponding to zeros or poles of the S-matrix. $S(p_0;0,1)$ has a double pole at $p_{0} = i$ ($E=0$) where
\eq{\psi_{+}(x) =  4e^{-\frac{x}{2}} \left(c_0-c_{+}\right) K_1(2e^{-\frac{x}{2}}), \qquad \psi_{0}(x) = 4 e^{-\frac{x}{2}} \left(c_+-c_{0}\right) K_0\left(2 e^{-\frac{x}{2}}\right).}
As shown in figure \ref{Plot1}, $\psi_{0}$ is a bound state. Although $\psi_{+}$ is not normalizable, it asymptotes to a constant so its momentum is localized at $p_{1}=0$. It corresponds to an anomalous threshold where the relative separation between the particles stays fixed. The presence of an anomalous threshold is usually associated with a double pole in the S-matrix \cite{Coleman:1978kk}, which is indeed the case here. Other singularities of the S-matrix are simple zeros and simple poles but they have negative energy hence are not physical.

For $s_A=2$, the first-order Darboux transform is not sufficient to separate the equations, and one needs to consider the second-order Darboux transform of the form $Q = (d + A_2) (d + A_1)$. However, solving for the unknowns is still nontrivial and the difficulty increases with the order. One may hope to use the bootstrap method to obtain the S-matrix for $s_A=2$ from the known S-matrices for lower spins by letting one particle go to its pole. Yet we have shown that the only physical pole for $s_A=1$ corresponds to an anomalous threshold so does not form a bound state.

\subsubsection{The compact $\mathfrak{su}(2)$ model}
Although the direct solution of the matrix Schr\"odinger problem for general spins is beyond reach for now, the solutions we found for spins 0, $1/2$ and 1 are sufficient for the fermionic model, where the spins are in the fundamental representation of $\mathfrak{su}(2)$. There is again an auxiliary $\mathfrak{su}(2)$ symmetry that hops spins from one site to another. The total number of spins gives us another $\mathfrak{u}(1)_N$ that commutes with both the global $\mathfrak{su}(2)_G$ and the auxiliary $\mathfrak{su}(2)_A$ symmetries. Because the vector space is finite-dimensional, we can write out the bases of two-particle states explicitly. They are organized into multiplets carrying charges under the symmetries, as shown in table \ref{bases}.
\begin{table}[htbp]
\begin{center}
    \begin{tabular}{c c c c}
        \hline
        $\mathfrak{u}(1)_{N}$ & $\mathfrak{su}(2)_{G}$ & $\mathfrak{su}(2)_{A}$ & States\\ \hline\hline \\
        0 & 0 & 0 & $|0,0\rangle$  \\ \\
        1 & $\dfrac{1}{2}$ & $\dfrac{1}{2}$ & $\mid\uparrow,0\rangle$,\quad $|0,\uparrow\rangle$,\quad $\mid\downarrow,0\rangle$,\quad $|0,\downarrow\rangle$\\ \\ 
        2 & 1 & 0 & $\mid\uparrow, \uparrow\rangle$,\quad $\mid\uparrow, \downarrow\rangle\, + \mid\downarrow, \uparrow\rangle$,\quad $\mid\downarrow, \downarrow\rangle$ \\ \\
        2  & 0 & 1 & $\mid\uparrow\downarrow,0\rangle$, \quad $\mid\uparrow, \downarrow\rangle\, - \mid\downarrow, \uparrow\rangle$,\quad $|0,\uparrow\downarrow\rangle$  \\ \\
        3 & $\dfrac{1}{2}$  & $\dfrac{1}{2}$ & $\mid\uparrow\downarrow, \uparrow\rangle$,\quad $\mid \uparrow,\uparrow\downarrow\rangle$,\quad $\mid\uparrow\downarrow, \downarrow\rangle$,\quad $\mid \downarrow,\uparrow\downarrow\rangle$ \\ \\
        4 & 0 & 0 & $\mid\uparrow\downarrow,\uparrow\downarrow\rangle$ \\ \\ \hline
     \end{tabular}
  \end{center}
    \caption{Bases of two-particle states in the fermionic Hubbard-Toda model.}
    \label{bases}
\end{table}

The two-body S-matrix decomposes into total spin $s_G=0$ and $s_G=1$ components as 
\eq{\mathbb{S}(p) = S_{s_G=0}(p) ~\!\mathcal{P}_0  + S_{s_G=1}(p) ~\!\mathcal{P}_1.}
We will focus on the case when there are $N=2$ spins between two sites and exactly one spin at each site $\Delta s \equiv N_1 - N_2= 0$. This corresponds to having exactly one spin at each site and is relevant to spin chains. Note that 
in the $N=2$ sector, $s_{G}=0$ corresponds to $s_{A} = 1$ and vice versa.
We may read off each component from the results in the previous section.
\EQ{\mathbb{S}(p) &= S(p;\Delta s=0,s_{A}=1)~\!\mathcal{P}_{0} + S(p;\Delta s=0, s_{A}=0)~\!\mathcal{P}_{1} \\
&=\frac{\G(1+ip)}{\G(1-ip)}\left(\frac{p-i~\!\mathbb{P}}{p-i}\right).}
The operator in the bracket is nothing other than the fundamental R-matrix for the $\mathfrak{su}(2)$ spin chain. Diagonalizing it with the quantum inverse scattering method described in the previous section, we obtain a set of Bethe ansatz equations for the fermionic $\mathfrak{su}(2)$ Hubbard-Toda model
\EQ{
e^{ip_{k}L} &=\prod^{K}_{\ell\ne k}\frac{\G\left(1+i(p_{k}-p_{\ell})\right)}{\G\left(1-i(p_{k}-p_{\ell})\right)}\prod^{M}_{i=1} \frac{p_{k}-\l_{i}-\frac{i}{2}}{p_{k}-\l_{i}+\frac{i}{2}} \\
\prod^{K}_{k=1}\frac{\l_{i} - p_{k} + \frac{i}{2}}{\l_{i} - p_{k} - \frac{i}{2}} &= e^{2\pi i\t}\prod^{M}_{j\ne i}\frac{\l_{i}-\l_{j} + i}{\l_{i}-\l_{j} - i}.}

\section{Solution by the Bethe/Gauge correspondence}\label{section4}
The quantum inverse scattering method that we discussed in the
previous section allows us to diagonalize the integrable system by
finding the Bethe ansatz equations. The method is standard albeit
somewhat technical. 
There is a novel way to obtain the Bethe ansatz equations 
directly from the corresponding gauge theory discovered by Nekrasov
and Shatashvili. The so-called Bethe/Gauge correspondence relates the 
supersymmetric vacua of $\mathcal{N}=2$ field theories with the 
eigenstates of the quantum integrable systems. 

Consider an ${\cal N}=2$ supersymmetric gauge theory in four
dimensions with gauge group $G$ of rank $r$. As above, the low-energy
physics on the Coulomb
branch is determined by a holomorphic curve $\Sigma$ of genus $r$ and
a meromorphic differential $\lambda_{\rm SW}$. Picking a canonical set of basis
cycles $\{A_{I},B_{I}\}$, with $A_{I}\cap B_{J}=\delta_{IJ}$
for $I,J=1,\ldots, r$, we have 
\eq{ \vec{a} \,\,=\,\, 
\frac{1}{2\pi i} \oint_{\vec{A}}\, \lambda_{\rm
  SW}, \qquad 
\vec{a}^{D} \,\,=\,\,\frac{1}{2\pi i}\, \frac{\partial
  \mathcal{F}}{\partial \vec{a}}
 \,\,=\,\,
\frac{1}{2\pi i} \oint_{\vec{B}}\, \lambda_{\rm
  SW}, \label{SWl} }
where $\vec{a}=(a_{1},\ldots,a_{r})$ with similar notation for other
$r$-component vectors. These relations determine the prepotential
$\mathcal{F}=\mathcal{F}(\vec{a})$ which in turn determines the exact
low-energy effective action on the Coulomb branch. As discussed in
section \ref{CxIntSys} above, the moduli of the Seiberg-Witten curve $\Sigma$ can
be identified with the Poisson-commuting Hamiltonians of a complex
classical integrable system. Further the periods of $\lambda_{\rm SW}$
around half of the basis cycles, for example the cycles
$\{B_{I}\}$, $I=1,\ldots,r$, correspond to canonical action
variables for the complex integrable system. 
To define a real quantum integrable system we need to choose a
middle-dimensional real slice of the Coulomb branch and then quantize
the corresponding action variables appropriately. In the Bethe/Gauge
correspondence, these two steps are accomplished simultaneously by
introducing an $\Omega$-background in one plane of the
4d spacetime of the gauge theory. More precisely we
consider the corresponding Nekrasov partition function \cite{Nekrasov:2002qd, Nekrasov:2003rj}    
\eq{\mathcal{Z}(\vec{a}, \epsilon_{1}, \epsilon_{2}), }
with deformation parameters $\epsilon_{1}$, $\epsilon_{2}$. 
Taking the limit $\epsilon_{2}\rightarrow 0$ with
$\epsilon=\epsilon_{1}$ held fixed, we define
a ``quantum'' prepotential
\eq{
\mathcal{F}  \left(\vec{a}, \epsilon\right) = \lim_{\epsilon_{2}
\rightarrow 0}\,
\epsilon_{1}\epsilon_{2}\, \log\,
\mathcal{Z}(\vec{a}, \epsilon_{1},
\epsilon_{2})\Big|_{\epsilon_{1}=\epsilon}\ .
 }
which reduces to the prepotential of the undeformed theory in the limit 
$\epsilon\rightarrow 0$. 

In the presence of the $\Omega$-background in one plane,
4d Lorentz invariance is broken and one obtains a
2d effective theory with $\mathcal{N}=(2,2)$
supersymmetry in the orthogonal plane. The supersymmetric vacua of
this theory are determined by the stationary points of an 
effective superpotential
\eq{
\mathcal{W}\left(\vec{a},\epsilon\right) = \frac{1}{\epsilon}\,
\mathcal{F}\left(\vec{a},\epsilon\right) \,\,-\,\,2\pi i\,
\vec{k}\cdot
\vec{a},
\label{wa} }
where the vector of integers, $\vec{k}\in\mathbb{Z}^{r}$ 
corresponds to a choice of branch for the perturbative 
logarithms appearing in $\cal{F}$. In the following we will suppress
the second term in (\ref{wa}) and work instead with a multi-valued
superpotential. 
The stationary points of this
potential correspond to states in the spectrum of a quantum integrable
system where the deformation parameter is identified with Planck's
constant as $\epsilon=-i\hbar$. 
In particular, working to leading order in $\epsilon$ we
obtain the quantization condition   
\eq{
\frac{\partial \mathcal{W}}{\partial \vec{a}}\,\,=\,\,0 \quad \Rightarrow \quad
\vec{a}^{D}\,\,\,\in\,\, \epsilon\,\mathbb{Z}^{r}. \label{q1} }
In particular, setting $\epsilon=-i\hbar$, this relation imposes the
condition ${\rm Re}\,\vec{a}^{D}=0$, which picks out a
middle-dimensional real slice of the Coulomb branch. Then 
${\rm Im}\,\vec{a}^{D}$ correspond to the canonical action variables of
the corresponding real integrable system and (\ref{q1}) coincides
with the Bohr-Sommerfeld quantization condition for this
system. Higher-order corrections in $\epsilon$ correct the
Bohr-Sommerfeld condition to give an exact quantization of the system.  
Values of the conserved Hamiltonians in each quantum state are
determined by the resulting on-shell values of the superpotential.  

An important feature of integrable systems captured by 
the Bethe/Gauge correspondence is that a single complex integrable
system can give rise to several inequivalent real integrable systems. 
In supersymmetric gauge theory this feature is related to the 
electro-magnetic duality group of the low-energy theory on the Coulomb
branch which corresponds to the group $Sp(2r,\mathbb{Z})$ of modular
transformations of the Seiberg-Witten curve $\Sigma$. 
The basis cycles defined above 
transform linearly under the action of the modular group, giving a new
set of quantization conditions. For example, performing a
$\mathbb{Z}_{2}$ electro-magnetic duality transformation, we obtain
the dual superpotential 
\eq{
\mathcal{W}_{D}\left(\vec{a}^{D}\right) =
\mathcal{W}\left(\vec{a}\right)\,\,+ \,\,\frac{2\pi i}{\epsilon}
\,\vec{a}\cdot\vec{a}^{D}, \label{leg2} }
whose F-term equations give rise to dual quantization conditions
\eq{
\frac{\partial \mathcal{W}_{D}}{\partial
\vec{a}^{D}}\,\,=\,\,0  \quad \Rightarrow \quad
\vec{a}\,\,\in\,\,\epsilon\,\mathbb{Z}^{r}. \label{q2} }
In the classical limit, $\epsilon\rightarrow0$, these give  
the reality condition ${\rm Re}\,\vec{a}=0$, which yields a 
real integrable system inequivalent to the one discussed
above. Working at non-zero $\epsilon=-i\hbar$ provides a quantization
of this system.  In the terminology of \cite{Nekrasov:2009rc} this is known as the 
B-quantization, while the original condition (\ref{q1}) corresponds to the
A-quantization. More 
generally, each element of the low-energy duality group yields a
distinct quantum integrable system in this way.       

We now turn to the integrable systems corresponding to elliptic 
quiver gauge theories. For simplicity, we will focus on the $N=2$
case, in other words 
the $\hat{A}_{1}$ quiver with gauge group $G=U(1)\times
SU(K)_{1}\times SU(K)_{2}$,
whose IIA brane construction is illustrated in figure \ref{brane}. As above we
have Coulomb branch
parameters $a^{(1)}_{k}$ and $a^{(2)}_{k}$ for the two $SU(K)$ factors
in $G$ . 
The corresponding cycles on the Seiberg-Witten curve are $\vec{A}^{(\alpha)}$ for
$\alpha=1,2$. We also define complexified couplings
$\t_\a$ for $SU(K)_{\alpha}$ and bi-fundamental masses
$m_{\alpha}$ for $\alpha=1,2$. 
For cosmetic reasons, the Coulomb branch parameters 
$a^{(1)}_{k}$ and $a^{(2)}_{k}$ 
will be renamed $a_{k}$ and $b_{k}$ respectively for $k=1,\ldots,K$
and will be organized as $K$-component vectors 
$\vec{a}=(a_{1},\ldots,a_{K})$, $\vec{b}=(b_{1},\ldots,b_{K})$. 
The solution of the model is then specified by the
quantum prepotential $\hat{\mathcal{F}}(\vec{a}, \vec{b}, \epsilon)$.

The $N=2$ case of the $K$-body elliptic spin Calogero-Moser model,  
considered above as a complex classical integrable system, corresponds
to the special strong-coupling point $\t_1=0$ of the
$\hat{A}_{1}$ quiver theory. According to the discussion above, to
choose a real integrable system and quantize it, we need to select a
set of basis cycles for the curve. To begin we change basis to cycles 
$\vec{A}_{+}=\vec{A}^{(1)}$ and 
$ \vec{A}_{-}=\vec{A}^{(1)}-\vec{A}^{(2)}$,
with similar definitions for $\vec B_{\pm}$. The corresponding periods are 
\eq{
 \vec{a}_{\pm} \,\,=\,\, 
\frac{1}{2\pi i} \oint_{\vec{A}_{\pm}} \lambda_{\rm
  SW}, \qquad 
\vec{a}^{D}_{\pm} \,\,=\,\,\frac{1}{2\pi i}\, \frac{\partial
  \mathcal{F}}{\partial \vec{a}_{\pm}}
 \,\,=\,\,
\frac{1}{2\pi i} \oint_{\vec{B}_{\pm}}\lambda_{\rm
  SW} ,   \label{SWl} 
}
which are related to the original Coulomb branch variables via
\eq{ 
\vec{a}=\vec{a}_{+}, \qquad 
\vec{b}=\vec{a}_{+}-\vec{a}_{-} .
\label{ab}
}
We will shortly see that the appropriate semi-classical quantization
condition for the real elliptic Calogero-Moser model is 
\eq{ \vec{a}_{+}^{D}\in \e\, \mathbb{Z}^{K}, \qquad 
\vec{a}_{-}\in \e\, \mathbb{Z}^{K}, }
which corresponds to a mixed scheme in which we choose the
``A-quantization'' for the dynamical 
variables associated with the diagonal $SU(K)$ and the
``B-quantization'' for those associated with the off-diagonal
$SU(K)$. Accordingly we define a multi-valued effective
superpotential 
\eq{\mathcal{W}\left(\vec{a}_{+},\vec{a}_{-}^{D}\right) 
=\frac{1}{\epsilon}\left[{\mathcal{F}}
\left(\vec{a}_{+},\vec{a}_{-}\right)\,\,-\,\, 
2\pi i \,\vec{a}_{-}\cdot \vec{a}_{-}^{D}\right],
\label{spAB}
}
where 
\eq{  
\mathcal{F}\left(\vec{a}_{+},\vec{a}_{-}\right)  =
\hat{\mathcal{F}}\left(\vec{a}_{+}, 
\vec{a}_{+}-\vec{a}_{-},\epsilon\right).  }
In principle, the on-shell values of the superpotential (\ref{spAB})
should provide a quantization of the elliptic spin Calogero-Moser
model for all values of the parameters. In this paper, we are
primarily interested in the large-volume limit ${\rm
  Im}\,\tau\rightarrow \infty$ with $\t_1$ held fixed where 
we can compare with the solution of
the model obtained using the asymptotic Bethe ansatz in the preceding
sections. In the quiver gauge theory, this limit has a clear
interpretation: as ${\rm
  Im}\,\t_2\rightarrow \infty$, 
the factor $SU(K)_{2}$ in the gauge group is frozen out becoming an
$SU(K)$ flavor symmetry. The resulting theory is an $A_{1}$ quiver,
in other words an $SU(K)_{1}$ gauge theory with $K$ hypermultiplets in the
fundamental representation and $K$ hypermultiplets in the
anti-fundamental representation. We denote the corresponding hypermultiplet 
masses $\vec{m}_{\rm F}$ and 
$\vec{m}_{\rm AF}$ respectively. This theory has fixed gauge coupling 
$\t_1$ and Coulomb branch parameters $\vec{a}=\vec{a}^{(1)}$. 
In the absence of the $\Omega$-deformation, we can
identify the mass parameters of this theory as follows:
$\vec{m}_{\rm F}= \vec{b}$ and 
$\vec{m}_{\rm AF}= \vec{b}+\vec{m}$ where $\vec{m}=(m,\ldots,m)$
with $m=m_1+m_2$ as above. In the following, we will propose
that these identifications are corrected slightly for non-zero
$\epsilon$ 
to read 
\eq{
\vec{a}=\vec{a}^{(1)}+\frac{3}{2}\vec{\epsilon},   \qquad
\vec{m}_{\rm F}= \vec{b}+\frac{3}{2}\vec{\epsilon},
  \qquad \vec{m}_{\rm AF}=
 \vec{b}+\vec{m}-\frac{1}{2}\vec{\epsilon},\label{e2}
}
with $\vec{\epsilon}=(\epsilon,\ldots,\epsilon)$. 

In the limit ${\rm Im}\,\t\rightarrow \infty$ 
the quantum prepotential of the affine quiver theory goes
over to that of the linear quiver described in the preceding
paragraph with prepotential denoted $\mathcal{F}_{\rm linear}(\vec{a},
\vec{m}_{\rm F}, \vec{m}_{\rm AF})$.   
Working at large but finite ${\rm Im}\,\t$ corresponds to weakly
gauging the $SU(K)$ flavor symmetry, and we restore the leading weak-coupling dynamics of the $SU(K)_{2}$ vector multiplet. Thus we have
two contributions to the prepotential 
$\hat{\mathcal{F}}(\vec{a},\vec{b},\epsilon)=
\hat{\mathcal{F}}_{1}+\hat{\mathcal{F}}_{2}$ where 
\eq{
\hat{\mathcal{F}}_{1}  = \mathcal{F}_{\rm linear}(\vec{a},
\vec{m}_{\rm F}, \vec{m}_{\rm AF}),    
}
with the parameter identifications described above and 
\eq{ 
\hat{\mathcal{F}}_{2}(\vec{b}) =  -2\pi i \t\sum_{k=1}^{K} \frac{b_{k}^{2}}{2} \,-\,
\e \sum_{k,\ell=1}^{K}\, \omega_{\e}\left(b_{k}-b_{\ell}\right),
\label{fhat2}
} 
where $\omega_{\e}$ satisfies $\omega'_{\e}(x) = -\log
\G(1+x/\e)$. The first term on the right-hand side is a classical
contribution while the second corresponds to the one-loop contribution
of the $SU(K)_{2}$ vector multiplet. 

Including both the above contributions to $\hat{\mathcal{F}}$ we form the
effective superpotential $\mathcal{W}$ using (\ref{spAB}).  
According to our chosen quantization scheme, the above superpotential
should be evaluated on shell at the quantized values  
\eq{
\vec{a}_{-} = \vec{a}-\vec{b}=-\vec{n}\epsilon, \qquad \vec{n}\in \mathbb{Z}^{K}, \label{e3} }
and then stationarized with respect to $\vec{a}_{+}$. In terms of the
parameters of the linear quiver, the quantization of 
$\vec{a}_{-}$ corresponds to selecting the
values $\vec{a}=\vec{m}_{\rm F}-\vec{n}\epsilon$. 

The gauge theory relevant for describing the spin Calogero-Moser model with unbroken 
$\mathfrak{sl}(2, \mathbb{R})$ symmetry is the $\hat{A}_{1}$ quiver
theory in the limit of infinite coupling for the off-diagonal 
gauge coupling $\t_1\rightarrow 0$. On the other hand, the 
Nekrasov partition function for the quiver theory and the corresponding
superpotential $\mathcal{W}_{1}$ is defined as a series in
powers of the instanton factor $q_{1}=\exp(2\pi i\t_1)$ which
becomes of order one near this point. Remarkably we
can pass to a dual description of the theory which effectively resums
the instanton series and allows us to obtain explicit results in the
limit $\t_1\rightarrow 0$. The dual description corresponds to
the world-sheet theory of vortex strings in the four-dimensional gauge
theory. In particular, the results of \cite{Dorey:2011pa} allow us to evaluate 
the difference 
\eq{ 
\Delta{\mathcal{F}} =   
\mathcal{F}_{\rm linear}(\vec{a},
\vec{m}_{\rm F}, \vec{m}_{\rm AF})\Big|_
{\vec{a}=\vec{m}_{\rm F}-\vec{n}\epsilon}
\,\,\,-\,\,\,
\mathcal{F}_{\rm linear}(\vec{a},
\vec{m}_{\rm F}, \vec{m}_{\rm AF})\Big|_
{\vec{a}=\vec{m}_{\rm F}} 
}
between the value of the quantum prepotential $\mathcal{F}_{\rm
linear}$ of the linear quiver at
the on-shell value $\vec{a}=\vec{m}_{\rm F}-\vec{n}\epsilon$, where  
$\vec{n}=(n_{1},\ldots,n_{K})$ is a vector of non-negative integers,
and its value at the root of the Higgs branch $\vec{a}=\vec{m}_{\rm
  F}$. The duality of \cite{Dorey:2011pa} equates $\Delta{\mathcal{F}}/\epsilon$ to the on-shell value of the superpotential $\mathcal{W}^{\rm 2d}$ 
of the vortex world-volume theory. 
The latter is a function of $M$ complex
variables $\sigma_{i}$ with $i=1,\ldots,M$ corresponding to the
scalars in the twisted chiral multiplets of the 2d theory. Explicitly
we have \cite{Witten:1993yc}
\eq{ 
\CW^\text{2d}  =  2\pi i\hat{\tau}
\sum^{M}_{i=1}\s_{i} + 
\sum_{i,j=1}^{M}
f(\s_{i} -\s_{j}-\e) + \sum^{M}_{i=1}\sum^{K}_{k=1} 
f(\s_i - \tilde{M}_k)-\sum^{M}_{i=1}\sum^{K}_{k=1}f(\s_i - M_k), 
\label{w2d} } 
where $f(x) = x \log (x/\e) - x$. The parameters
appearing in the superpotential correspond to the masses 
$\vec{M}_{\rm F}=(M_{1},\ldots,M_{K})$ and $\vec{M}_{\rm AF}=
(\tilde{M}_{1},\ldots,\tilde{M}_{K})$ of chiral multiplets in the
fundamental and the anti-fundamental of the 2d gauge group $U(M)$ and the 
complexified Fayet-Iliopoulos parameter $\hat{\tau}$ of the vortex theory. These
are related to the parameters of the 4d linear quiver theory via 
\eq{
\hat{\tau}=\tau + \frac{1}{2}(M+1), \qquad  \vec{M}_{F} = \vec{m}_{\rm F} - \frac{3}{2} \vec{\e} \ ,
  \qquad  
  \vec{{M}}_{{AF}} = \vec{{m}}_{{\rm AF}} + \frac{1}{2} \vec{\e}\ .
\label{ident}
}
In order to 
determine the on-shell value of the 2d superpotential we compute its
stationary values with respect to the 2d fields $\s_{i}$, which yields the
Bethe ansatz-like equations 
\eq{
\prod_{k=1}^{K}
\frac{\s_{i}-M_{k}}{\s_{i}-\tilde{M}_{k}} =
e^{2\pi i \t_1}
\prod_{j \neq i}^{M} \frac{\s_{i}-\s_{j} -\epsilon}
{\s_{i}-\s_{j} +\epsilon} \label{baex}. }

To evaluate the superpotential (\ref{spAB}) at the on-shell values of 
$\vec{a}_{-}$, we include all the contributions described above: 
\eq{ 
\mathcal{W}(\vec{a}_{+}) 
=  \CW^\text{2d}+ \frac{1}{\epsilon}
\mathcal{F}_{\rm linear}(\vec{a},
\vec{m}_{\rm F}, \vec{m}_{\rm AF})\Big|_
{\vec{a}=\vec{m}_{\rm F}} + \frac{1}{\epsilon}\hat{\mathcal{F}}_{2}.  }
To complete our calculation we need to evaluate the final term which
corresponds to the value of the superpotential at the Higgs branch
root. Here we will use the fact that the theory at the root reduces to 
that of the weakly-gauged $SU(K)_{2}$ vector multiplet coupled to a
single adjoint hypermultiplet of mass $m=m_1+m_2$. The
classical prepotential of this theory is already accounted for in 
$\hat{\mathcal{F}}_{2}$ as is the one-loop contribution of the vector
multiplet. The remaining contribution is that of the adjoint
hypermultiplet. Thus we must have 
\eq{ 
\mathcal{F}_{\rm linear}(\vec{a}, \vec{m}_{\rm F}, \vec{m}_{\rm AF})\Big|_{\vec{a}=\vec{m}_{\rm F}}=       
\e\sum_{k,\ell=1}^{K}\, \omega_{\e}\left(b_{k}-b_{\ell}+m\right) 
.} 
This equality can also be understood directly from the IIA string
theory construction of the duality of \cite{Dorey:2011pa}. 

To obtain the superpotential as a function of $\vec{a}_{+}$, we impose the equations
of motion (\ref{baex}) for $\sigma_{j}$ and eliminate $\vec{M}_{\rm F}$,
$\vec{M}_{\rm AF}$ and $\vec{b}$ in terms of $\vec{a}_{+}$ using 
(\ref{ab}, \ref{e2}, \ref{e3}, \ref{ident}).  
Finally we minimize the resulting superpotential with respect to
$\vec{a}_{+}$ to obtain 
\EQ{
e^{\frac{b_{k}}{\e}2\pi i \tau} &=  
\prod^{K}_{\ell=1}
\frac{\G\left(1+\frac{m}{\e} - \frac{b_{k\ell}}{\e}
  \right)\G\left(1+ \frac{b_{k\ell}}{\e}
  \right)}{\G\left(1+\frac{m}{\e} + \frac{b_{k\ell}}{\e} \right)
\G\left(1-\frac{b_{k\ell}}{\e} \right)}
\, \prod^{M}_{i=1}
\frac{b_{k} - \s_{i}}{b_{k} - \s_{i} - m} 
\\
 \prod^{K}_{k=1}\frac{\s_{i} - b_k}{\s_{i} - b_{k}+m} &= 
e^{2\pi i\t_1}\prod^{M}_{j \ne i}\frac{\s_{i}-\s_{j} - \e}
{\s_{i}-\s_{j}+\e},
 }
where $b_{k\ell}=b_k-b_\ell$ and, for on-shell values of $\vec{a}_{-}$ we
have 
\eq{
\vec{b} = \vec{a}_{+}+\vec{n}\epsilon . 
}
Once we identify the twisted chiral scalars $\s_i$ with the magnon
rapidities $\l_i$ as $\s_i = \l_i -m/2$, these are precisely the
Bethe ansatz equations for the spin Calogero-Sutherland model
(\ref{BAEsCM1}, \ref{BAEsCM2})! Here we identify the Coulomb branch
parameter $b_{k}$ with the particle momentum $p_{k}$ and the mass of
the adjoint hypermultiplet $m$ with $2s\e $ where $s\in \mathbb{Z}/2$ is the
spin of the $\mathfrak{sl}(2,\mathbb{R})$ representation at each
site. As above we set $\epsilon=-i\hbar$. In fact these equations hold
for both $\mathfrak{su}(2)$ representations with $s>0$ and 
$\mathfrak{sl}(2,\mathbb{R})$ representations corresponding to $s<0$. 

Now we turn to the Hubbard-Toda chain. Again we will focus on the
$N=2$ case where the spins lie in lowest-weight representations of 
$\mathfrak{sl}(2,\mathbb{R})$. To find an exact quantization of this
system using the Bethe/Gauge correspondence we will start with the
elliptic $\hat{A}_{2}$ quiver with gauge group 
$G=U(1)\times SU(K)_{1}\times SU(K)_{2}\times SU(K)_{3}$,
whose IIA brane construction is illustrated in figure \ref{Inozemtsev}. As above we
have Coulomb branch
parameters $a^{(1)}_{k}$, $a^{(2)}_{k}$ and $a^{(3)}_{k}$ 
for the three $SU(K)$ factors
in $G$ with $k=1,\ldots,K$. 
The corresponding cycles on the Seiberg-Witten curve are $\vec{A}^{(\alpha)}$ for
$\alpha=1,2,3$. We also define complexified couplings
$\t_\a$ for $SU(K)_{\alpha}$ and bi-fundamental masses
$m_{\alpha}$ for $\alpha=1,2,3$. 
For cosmetic reasons, the Coulomb branch parameters 
$a^{(1)}_{k}$, $a^{(2)}_{k}$ and $a^{(3)}_{k}$
will be renamed $c_{k}$, $a_{k}$ and $b_{k}$ respectively for $k=1,\ldots,K$
and will be organized as $K$-component vectors 
$\vec{c}$, $\vec{b}$ and $\vec{a}$. 
The solution of the model is then specified by the
quantum prepotential $\hat{\mathcal{F}}(\vec{a}, \vec{b}, \vec{c},\epsilon)$.

To begin we change basis to cycles 
\eq{ 
\vec{A}_{+}=\vec{A}^{(1)},   \qquad \qquad  
\vec{A}_{-}=\vec{A}^{(2)}-\vec{A}^{(1)},  \qquad \qquad  
\vec{A}'_{-}=\vec{A}^{(3)}-\vec{A}^{(2)},  
\label{cyc} 
}with appropriate definitions for the conjugate cycles 
$\vec B_{+}$, $\vec B_{-}$ and $\vec B_{-}'$ which we will not need here. 
The corresponding periods are
\eq{
\vec{a}_{+}=\vec{a},\qquad\qquad 
\vec{a}_{-}=\vec{b}-\vec{a}, \qquad \qquad \vec{a}'_{-} =
\vec{c}-\vec{b}+\vec{m}, 
\label{cyc23}
}  
where $\vec{m}=(m,\ldots,m)$ and $m=m_1+m_2+m_3$.
The inverse relations are  
\eq{
\vec{a} =\vec{a}_{+}, \qquad\qquad \vec{b} =
\vec{a}_{+}+\vec{a}_{-}, \qquad\qquad
\vec{c}  = \vec{a}_{+}+\vec{a}_{-}+\vec{a}'_{-}-\vec{m} .
\label{abc}
}
 
As in the previous example we will use a mixed quantization scheme
where the diagonal cycles are treated in the A-quantization and the
off-diagonal ones are treated using the B-quantization. Thus the
quantization conditions take the form 
\eq{ \vec{a}_{+}^{D}\in \e\, \mathbb{Z}^{K}, \qquad \qquad  
\vec{a}_{-}, \,\,\vec{a}_{-}'\in \e\, \mathbb{Z}^{K},  } 
with the corresponding superpotential 
\eq{ 
\mathcal{W}  \left(\vec{a}_{+},{\vec{a}}_{-}^{D},{{\vec{a}}}'^{D}_{-}\right)  = 
\frac{1}{\epsilon}\Big[{\mathcal{F}}
\left(\vec{a}_{+},\vec{a}_{-},{{\vec{a}}'}_{-}\right)\,-\, 
2\pi i\, \vec{a}_{-}\cdot {\vec{a}}_{-}^{D}\,-\,
2\pi i\, {\vec{a}}'_{-}\cdot {{\vec{a}}}'^{D}_{-}\Big], 
\label{spABcc}
}
where
$
\mathcal{F}\left(\vec{a}_{+},\vec{a}_{-}, \vec a_{-}'\right) =
\hat{\mathcal{F}}(\vec{a}, \vec{b}, \vec{c}, \epsilon)$
and $\vec{a}$, $\vec{b}$, $\vec{c}$ are given by (\ref{abc}) above. 

As for the spin Calogero-Moser model we will focus on the weak-coupling
limit ${\rm Im}\,\t\rightarrow \infty$, where
$\tau=\t_1+\t_2
+\t_3$, holding ${\rm Im}\,\t_2$ fixed. For convenience we also set 
$\t_1=\t_3=\tau'$. In the resulting limit ${\rm Im}\,\t'$
becomes large so that the gauge group
factors $SU(K)_{1}$ and $SU(K)_{3}$ are weakly coupled. Once again
this limit leads to an $A_{1}$ linear quiver with gauge group
$SU(K)_{1}$ where $SU(K)_{2}$ and
$SU(K)_{3}$ are weakly-gauged flavor symmetries. 

In the absence of an
$\Omega$-deformation, the Coulomb branch vacuum expectation values of the linear quiver are 
$\vec{a}=\vec{a}^{(1)}$ while the 
fundamental and the anti-fundamental hypermultiplets have masses
$\vec{m}_{\rm F}=\vec{b}$, $\vec{m}_{\rm AF}=\vec{c}$.  
In the limit ${\rm Im}\,\t\rightarrow \infty$, 
the quantum prepotential of the affine quiver gauge theory goes
over to that of the linear quiver denoted $\mathcal{F}_{\rm linear}(\vec{a},
\vec{m}_{\rm F}, \vec{m}_{\rm AF})$ plus contributions from the weakly-gauged flavor symmetries which take the form 
\eq{
\mathcal{F}_{\rm weak} = \hat{\mathcal{F}}_{2}(\vec{b})+
\hat{\mathcal{F}}_{2}(\vec{c})+ 
\e\sum_{k,\ell=1}^{K}\, \omega_{\e}\left(b_{k}-c_{\ell}\right),
}
where $\hat{\mathcal{F}}_{2}(\vec{b})$, given in (\ref{fhat2}), 
is the classical and the one-loop vector multiplet contributions for
$SU(K)_{2}$ and  $\hat{\mathcal{F}}_{2}(\vec{c})$ is a similar term
for $SU(K)_{3}$. The final term represents the one-loop contribution
of the bi-fundamental hypermultiplet of $SU(K)_{2}\times SU(K)_{3}$. 
The resulting superpotential 
\eq{ 
\mathcal{W} = \frac{1}{\epsilon}\left(
\mathcal{F}_{\rm linear}+\mathcal{F}_{\rm
  weak}\right),
}
should be evaluated at the on-shell values of $\vec{a}_{-}$
and $\vec{a}'_{-}$ and then stationarized with respect to
$\vec{a}_{+}$. 

Once again we can
use the duality of \cite{Dorey:2011pa} to evaluate the contribution of the linear
quiver explicitly. Collecting the various contributions to the
superpotential and minimizing with respect to $\vec{a}_{+}$ yields the
following equations 
\EQ{ 
\left(e^{2\pi i\t}\right)^{\frac{b_{k}+c_{k}}{2\e}} 
&= 
\prod_{\ell=1}^{K} \frac{\Gamma\left(1+\frac{b_{k\ell}}{\e}\right)\Gamma\left(1+\frac{c_{k\ell}}{\e}\right)\Gamma\left(1+\frac{b_{k}-c_{\ell}}{\e}\right)\Gamma\left(1+\frac{m}{\e}+ \frac{c_{k}-b_{\ell}}{\e}\right)}
{\Gamma\left(1-\frac{b_{k\ell}}{\e}\right)\Gamma\left(1-\frac{c_{k\ell}}{\e}\right)\Gamma\left(1+\frac{b_{\ell}-c_{k}}{\e}\right)\Gamma\left(1+\frac{m}{\e}+
    \frac{c_{\ell}-b_{k}}{\e}\right)} 
\prod_{i=1}^{M} 
\frac{\s_{i}-b_{k}}{\s_{i}-c_{k}}  \\ 
   \prod_{k=1}^{K} 
\frac{\s_{i}-b_{k}}{\s_{i}-c_{k}} &=  
e^{2\pi i \t_2} \prod^{M}_{j \neq i} \frac{\s_i-\s_j-\e}{\s_i-\s_j+\e}.
\label{big} }
Now we take the Inozemtsev limit $\tau\rightarrow\infty$,
$m\rightarrow\infty$ with $\Lambda=m\exp(\pi i \tau/K)$ held fixed,
after which the first equation in (\ref{big}) reduces to  
\eq{ 
\left(\Lambda^{2K}\right)^{\frac{b_{k}+c_{k}}{2\e}} =
\prod_{\ell=1}^{K} \frac{\Gamma\left(1+\frac{b_{k\ell}}{\e}\right)\Gamma\left(1+\frac{c_{k\ell}}{\e}\right)\Gamma\left(1+\frac{b_{k}-c_{\ell}}{\e}\right)}
{\Gamma\left(1-\frac{b_{k\ell}}{\e}\right)\Gamma\left(1-\frac{c_{k\ell}}{\e}\right)\Gamma\left(1+\frac{b_{\ell}-c_{k}}{\e}\right)}
\prod_{i=1}^{M} 
\frac{\s_{i}-b_{k}}{\s_{i}-c_{k}}.} 
We can identify these equations with the Bethe ansatz equations of
the Hubbard-Toda chain by setting $b_{k}=p_{k}+s_{k}\e$,
$c_{k}=p_{k}-s_{k}\e$ where $\e=-i\hbar$ and $s_{k}\in \mathbb{Z}/{2}$
is the spin label of the $k$-th particle. The 2d fields $\s_{i}$
are identified with the magnon rapidities $\l_{i}$. The resulting equations
read 
\EQ{
e^{-iL\frac{p_{k}}{\hbar}} &= \prod _{\ell=1}^{K}
S_{0}\left(p_{k}-p_{\ell};s_{k},s_{\ell}\right)  \prod _{i=1}^{M} 
\frac{p_{k}-\l_{i}-is_{k}\hbar}{p_{k}-\l_{i}+is_{k}\hbar} \\ 
 \prod_{k=1}^{K} 
\frac{\l_{i}-p_{k}+is_{k}\hbar}{\l_{i}-p_{k}-
is_{k}\hbar} &=  
e^{2\pi i \t_2} \prod^{M}_{j \neq i}
\frac{\lambda_i-\l_j+i\hbar}
{\lambda_i-\l_j-i\hbar}.
 } 
Here we define the effective system size $L=-2K\log \Lambda$. The
central scattering phase is given as
\eq{S_{0}(p; s_{1}, s_{2}) = 
\frac{\G\left(1+\Delta s+\frac{p}{\e} \right)\G\left(1-\Delta
    s+\frac{p}{\e} \right)\G\left(-\bar s -
    \frac{p}{\e}\right)}{\G\left(1+\Delta s-\frac{p}{\e}
  \right)\G\left(1-\Delta s-\frac{p}{\e} \right)\G\left(-\bar s +
    \frac{p}{\e}\right)},}
with $\Delta s=s_{1}-s_{2}$ and $\bar{s}=s_{1}+s_{2}$. 
This gives a prediction for the 
scalar part of the S-matrix for the Hubbard-Toda model.

\acknowledgments
We thank Sungjay Lee for collaboration at an early stage of the project. We are indebted to Alexander Gorsky, Kazuo Hosomichi, Io Kawaguchi,
Ivan Kostov, Carlo Meneghelli, Nikita Nekrasov, Vasily Pestun, Didina Serban, Di Wang, Dan Xie, Masahito Yamazaki, Kentaroh Yoshida and Xinyu Zhang for very helpful discussions.

\appendix
\section{The Inozemtsev limit to the Hubbard-Toda model}\label{appendixA}

We can flow from the Calogero-Moser potential to the Toda potential by taking the Inozemtsev limit \cite{Inozemtsev1989finite} (see also \cite{D'Hoker:1998yh, Khastgir:1999pd, Chernyakov:2001wf}), i.e., sending the coupling and particle positions to infinity as
\eq{x_{k} = X_{k} + k\log \m^2, \qquad \Lambda^{K} = \m^{K}e^{2\pi i \t} \text{ fixed}.}
The Lax matrix for the inhomogeneous spin Calogero-Moser model is (\ref{gsCMLax})
\eq{L_{k\ell}(z) ]= \d_{k\ell}\left[p_{k} + \sum_{\a=1}^{N} S^{\a}_{kk}\z(z-z_{\a})\right]+(1-\d_{k\ell})\sum^{N}_{\a=1} S^\a_{k\ell}\, \frac{\s(x_{k\ell}+z-z_\a)}{\s(x_{k\ell})\s(z-z_\a)}~\!e^{x_{k\ell}(\psi(z)-\psi(z_\a))}.}
The Hubbard-Toda model arises from the inhomogeneous spin Calogero-Moser model by setting one inhomogeneity at half-period as $z_{N} = i\pi \t$ and setting the rest at the origin as $z_{\a} = 0$ for $\a = 1, \ldots, N-1$. We further set
\eq{Q^{N}_{k} = \tilde Q^{N}_k = \sqrt{\mu}, \qquad \sum^{N-1}_{\a=1}S^{\a}_{kk} = m - \mu.
}
For this configuration, the Lax matrix becomes 
\EQ{L_{k\ell}(z) &= \d_{k\ell}\left[p_{k} + \frac{m-\mu N}{N}\left(\z(z)-\z(z-\omega_{2})\right)\right]\\
&+ (1-\d_{k\ell})~\!e^{x_{k\ell}\!\!~\psi(z)}\left[A_{k\ell} \frac{\s(x_{k\ell}+z)}{\s(x_{k\ell})\s(z)}+\m \frac{\s(x_{k\ell}+z-\omega_{2})}{\s(x_{k\ell})\s(z-\omega_{2})}~\!e^{\z(\omega_{2})~\!\!x_{k\ell}}\right].}

We first examine the diagonal part. Using the asymptotic formula for $\z(z)$ as $\text{Im }\t \to \infty$,
one can show that in the Inozemtsev limit the diagonal part becomes
\eq{L_{kk}(z) = p_{k} + \frac{m-\mu N}{N}\left(\z(\omega_{2}) + \frac{1}{2}\coth\frac{z}{2}\right). \label{diagonal}
}
As we can shift the Lax matrix by a constant times the identity matrix without changing the spectral curve and hence the set of commuting Hamiltonians, we will absorb the site-independent constant in the diagonal part (\ref{diagonal}) by redefining the $v$ parameter in the spectral curve (\ref{spectralcurve}).

For the off-diagonal part, the overall factor $e^{x_{k\ell}~\!\!\psi(z)}$ can also be absorbed by a gauge transformation $L \mapsto gLg^{-1}$ of the diagonal form $g_{k\ell} = \delta_{k\ell}
~\!e^{x_k\psi(z)}$ that leaves the spectral curve invariant. We use an infinite-series representation \cite{Chernyakov:2001wf}
\eq{\frac{\s(x_{k\ell}+z)}{\s(x_{k\ell})\s(z)} = e^{\frac{\z(\omega_{1})}{\omega_{1}}x_{k\ell}z}\sum_{n \in \mathbb{Z}} \frac{e^{nz}}{1-e^{-2n~\!\!\omega_2-x_{k\ell}}}.}
In the $\text{Im}~\!\t \to \infty$ limit, the only non-zero contributions are from the $n \le 0$ terms in the summand. The $n=0$ term tends to 1 if $k > \ell$ and tends to 0 if $k < \ell$.  
It follows that the right-hand side is equal to $\sum_{n\leq0} e^{nz}$ when $k >\ell$ and is equal to $\sum_{n< 0} e^{nz}$  if $k < \ell$. 

In the variable $t = e^{z}$, the part of the Lax matrix that depends on the spin variables can be written more compactly as
\eq{L^\text{spin}_{k\ell}(z) = A_{k\ell}\left[\d_{k\ell}\frac{t+1}{2(t-1)}+\Theta_{k\ell}\frac{t}{t-1} + \Theta_{\ell k}\frac{1}{t-1}\right],}
where $\Theta_{k\ell}$ is the discrete Heaviside function taking value $1$ for $k > \ell$ and zero otherwise.
Using the Legendre relation 
\eq{\omega_{2}~\!\z(\omega_{1}) - \omega_{1}~\!\zeta(\omega_{2}) = \frac{i\pi}{2},}
and setting $\omega_{1} = i\pi$, $\omega_{2} =i\pi \t$, we may write
\eq{\frac{\s(x_{k\ell}+z-\omega_2)}{\s(x_{k\ell})\s(z-\omega_2)}~\!e^{\z(\omega_{2})~\!\!x_{k\ell}} = e^{\frac{\z(\omega_{1})}{\omega_{1}}x_{k\ell}z}\sum_{n \in \mathbb{Z}} \frac{e^{nz-n~\!\!\omega_2-\frac{x_{k\ell}}{2}}}{1-e^{-2n~\!\!\omega_2-x_{k\ell}}}.}
The dominant term in the Inozemtsev limit is $e^{X_{k}-X_{k-1}}$ for $n=0$ and $\Lambda^{K}t^{\pm 1}e^{\pm(X_{1} - X_{K})}$ for $n=\pm 1$.
For the part of the Lax matrix that depends on the dynamical variables $x_k, p_k$, the long-range interactions are exponentially suppressed and we obtain the nearest-neighbor interaction with the Toda potential
\EQ{L^\text{Toda}_{k\ell}(z) &= \d_{k\ell}~\!p_{k} + \d_{(k-1)\ell}~\! e^{\frac{X_{k-1}-X_{k}}2} + \d_{(k+1)\ell}~\! e^{\frac{X_{k}-X_{k+1}}2} \\
&+ \d_{k1}\d_{\ell K}\frac{\Lambda^{K}}{t}e^{\frac{X_{K}-X_{1}}2} + \d_{k K}\d_{\ell 1}\Lambda^{K}t~\!e^{\frac{X_{K}-X_{1}}2}.\label{TodaLax}}

\section{Proof of classical integrability}\label{appendixB}
Because the Hubbard-Toda model arises as a special limit from a classically integrable model, we expect integrability to persist.
In this appendix we prove this by showing that the Lax matrices are intertwined by the classical $r$-matrix of the Toda chain. Classical integrability relies on the existence of an $r$-matrix that intertwines the Lax matrix acting on two vector spaces
\eq{
\{L(z) \stackrel{\otimes}{,} L(w)\} = [r(z,w), L(z) \otimes \mathbb{I}] - [r^{*}(w,z), \mathbb{I} \otimes L(w)]. \label{rmatrix}
}
In index notation, $r = r_{km \ell n}~e_{k\ell} \otimes e_{mn}$ and $r^{*} = r_{\ell nkm}~e_{k\ell} \otimes e_{mn}$, where $e_{k\ell}$ is a matrix that has 1 in the $(k,\ell)$-th entry and 0 elsewhere. Let us define $r_{12} = r(z_{1},z_{2})\otimes \mathbb{I}$, $r_{23} = \mathbb{I} \otimes r(z_{2}, z_{3})$ and similarly for $r_{13}$. We also require that the $r$-matrices satisfy the classical Yang-Baxter equation
\eq{
[r_{12}, r_{13}] + [r_{12}, r_{23}] + [r_{13}, r_{23}] = 0,\label{cYB}
}
which appears as the semi-classical limit of the quantum Yang-Baxter equation (\ref{qYB}) if we write $\mathbb{R}(z) = 1 + \hbar\!~r(z) + \mathcal{O}(\hbar^{2})$.
One can see that (\ref{cYB}) implies the Jacobi identity for the tensor product Poisson bracket.
It follows from (\ref{rmatrix}) that the traces of powers of the Lax matrix are in involution: 
\EQ{\{L^{m}(z)\stackrel{\otimes}{,} L^{n}(w)\} &=\sum^{m-1}_{r=0}\sum^{n-1}_{s=0} L^{r}(z)\otimes L^{s}(w)\{L(z)\stackrel{\otimes}{,} L(w)\}L^{m-r-1}(z)\otimes L^{n-s-1}(w)\\
&= \sum^{m-1}_{r=0}\sum^{n-1}_{s=0} L^{r}(z)\otimes L^{s}(w)\big([r_{12}(z,w), L(z) \otimes \mathbb{I}] - [r_{21}(w,z), \mathbb{I} \otimes L(w)]\big) \\
&\qquad \qquad \ \times L^{m-r-1}(z)\otimes L^{n-s-1}(w) 
.
}
Taking the trace and using the cyclic property, we see that $\text{tr}~\!L^{n}(z)$ defines a one-parameter family of commuting conserved charges.

The $r$-matrix for the Toda chain is \cite{Jimbo:1985ua}
\eq{
r_{k\ell mn}(z,w) = -\left[\d_{k\ell} \frac{t+s}{2(t-s)} + \Theta_{k\ell}\frac{t}{t-s} +\Theta_{\ell k}\frac{s}{t-s}\right] \d_{kn}\d_{\ell m}\equiv-\D'_{k\ell}(z,w)\d_{kn}\d_{\ell m},} where $t=e^{z}, s=e^{w}$. Since it satisfies $r(z,w) = - r^{*}(w,z)$, we can simplify (\ref{rmatrix}) as
\eq{\{L(z) \stackrel{\otimes}{,} L(w)\} = [r(z,w), L(z) \otimes \mathbb{I} + \mathbb{I} \otimes L(w)].
\label{rmatrixsimp}}
Recall the Hubbard-Toda Lax matrix (\ref{LaxHT})
\eq{L_{k\ell}(z) = L^\text{Toda}_{k\ell}(z) + L^\text{spin}_{k\ell}(z),}
with $L^\text{spin}_{k\ell}(z) = A_{k\ell}~\!\D_{k\ell}(z)$ and
\eq{\D_{k\ell}(z) = \d_{k\ell}\frac{t+1}{2(t-1)} + \Theta_{k\ell}\frac{t}{t-1} + \Theta_{\ell k}\frac{1}{t-1}.}
We check that $L(z)$ and $r(z,w)$ satisfy (\ref{rmatrixsimp}). 
Because the position and momentum variables $x_{k}, p_{k}$ have trivial Poisson bracket with the spin variables $A_{k\ell}$, and it is known that $L^\text{Toda}(z)$ satisfies (\ref{rmatrixsimp}) with the Toda $r$-matrix, it suffices to check that $L^\text{spin}(z)$ also satisfies (\ref{rmatrixsimp}) with the same $r$-matrix. 
The left-hand side can be written as
\EQ{
\{L^\text{spin}(z)\stackrel{\otimes}{,} L^\text{spin}(w)\}_{k\ell mn} &\equiv \{L^\text{spin}_{km}(z), L^\text{spin}_{\ell n}(w)\}\\
&= (\d_{kn}A_{\ell m} - \d_{\ell m}A_{kn})\D_{km}(z)\D_{\ell n}(w).}
The right-hand side is (writing $r = r(z,w)$ for short and summing over the primed indices)
\EQ{
&[r(z,w), L^\text{spin}(z)\otimes \mathbb{I} + \mathbb{I} \otimes L^\text{spin}(w)]_{k\ell mn}\\ 
&= r_{k\ell m'n}L^\text{spin}_{m'm}(z)+r_{k\ell mm'}L^\text{spin}_{m'n}(w) - L^\text{spin}_{km'}(z) ~\! r_{m'\ell mn}-L^\text{spin}_{\ell m'}(w) ~\! r_{km'mn} \\
&= -\d_{kn}A_{\ell m}\left(\D'_{k\ell}\D_{\ell m}(z)-\D'_{km}\D_{\ell m}(w)\right) + \d_{\ell m}A_{kn}\left(\D'_{n\ell}\D_{k n}(z) - \D'_{k\ell}\D_{kn}(w)\right). \\
}
One can show that this matches the left-hand side by expanding out the equation above and rewriting it as a product of $\Delta(z)$ and $\Delta(w)$ using the identity
\eq{\frac{t+s}{2(t-s)}\frac{1}{s-1} - \frac{1}{t-s}\frac{t}{t-1} = \frac{t+1}{2(t-1)}\frac{1}{s-1}.}

\bibliographystyle{./utphys.bst}
\bibliography{./bib}
\end{document}